\documentclass[12pt]{article}

\usepackage{graphicx}

\def\gtwid{\mathrel{\raise.3ex\hbox{$>$\kern-.75em\lower1ex\hbox{$\sim$}}}}
\def\ltwid{\mathrel{\raise.3ex\hbox{$<$\kern-.75em\lower1ex\hbox{$\sim$}}}}
\def\square{\kern1pt\vbox{\hrule height 1.2pt\hbox{\vrule width 1.2pt\hskip 3pt
   \vbox{\vskip 6pt}\hskip 3pt\vrule width 0.6pt}\hrule height 0.6pt}\kern1pt}

\begin{document}

\begin{titlepage}

\begin{flushright}
UFIFT-QG-16-03
\end{flushright}

\vskip 1cm

\begin{center}
{\bf Precision Predictions for the Primordial Power Spectra \\
from $f(R)$ Models of Inflation}
\end{center}

\vskip 1cm

\begin{center}
D. J. Brooker$^{1*}$, S. D. Odintsov$^{2,3,4\star}$ and R. P. Woodard$^{1\dagger}$
\end{center}

\vskip .5cm

\begin{center}
\it{$^{1}$ Department of Physics, University of Florida,\\
Gainesville, FL 32611, UNITED STATES}
\end{center}

\begin{center}
\it{$^{2}$ Institute of Space Sciences (IEEC-CSIC) \\
C. Can Magrans s/n, 08193 Barcelona, SPAIN}
\end{center}

\begin{center}
\it{$^{3}$ ICREA, Passeig Lluis Companys 23, 08010 Barcelona, SPAIN}
\end{center}

\begin{center}
\it{$^{4}$ INFN Sez. di Napoli, Compl. Univ. di Monte S. Angelo, \\
Edificio G, Via Cinthia, I-80126, Napoli, ITALY}
\end{center}

\vspace{1cm}

\begin{center}
ABSTRACT
\end{center}
We study the power spectra of $f(R)$ inflation using a new technique in which 
the norm-squared of the mode functions is evolved. Our technique results in
excellent analytic approximations for how the spectra depend upon the function 
$f(R)$. Although the spectra are numerically the same in the Jordan and Einstein
frames for the same wave number $k$, they depend upon the geometries of these 
frames in quite different ways. For example, the power spectra in the two frames
are different functions of the number of e-foldings until end of inflation. We
discuss how future data on reheating can be used to distinguish $f(R)$ inflation 
from scalar-driven inflation.    

\begin{flushleft}
\end{flushleft}

\vskip .5cm

\begin{flushleft}
$^{*}$ e-mail: djbrooker@ufl.edu \\
$^{\star}$ e-mail: odintsov@ieec.uab.es \\
$^{\dagger}$ e-mail: woodard@phys.ufl.edu
\end{flushleft}

\end{titlepage}

\section{Introduction}

The proposal that the evolution of the universe is caused mainly by gravitation
attracts more and more attention. However, it has been realized that gravity
is not as simple as we thought and could be modified from standard General 
Relativity in several ways. Modified gravity theories are especially
attractive to explain the current phase of cosmic acceleration.

The first complete model of primordial inflation was the 1980 proposal by
Starobinsky to modify the gravitational Lagrangian by the addition of a
term quadratic in the Ricci scalar \cite{Starobinsky:1980te}. Although this
model was for decades eclipsed by scalar potential models, the increasingly 
tight bounds on the tensor-to-scalar ratio \cite{Ade:2015xua}, and the 
consequent elimination of the simplest potentials \cite{Ade:2015lrj}, have 
combined to produce a resurgence of interest in it \cite{Kehagias:2013mya}.

It has been realized lately that more general modifications of the Hilbert
Lagrangian, from $R$ to $f(R)$, may provide a consistent description of late 
time acceleration \cite{Capozziello:2002rd,Carroll:2003wy}, or even provide
a unified description of primordial inflation and dark energy 
\cite{Nojiri:2003ft}. A number of modified gravities which may consistently 
describe such a unified evolution of the universe are known 
\cite{Capozziello:2011et,Nojiri:2010wj}. $f(R)$ gravity has attracted the
main interest because it is ghost-free and reasonably simple.

It is quite remarkable that $f(R)$ gravity appears as a two-faced Janus: in
the Jordan frame it is a modified gravity theory, whereas it is a kind of 
scalar-tensor theory after conformal transformation to the Einstein frame.
The equivalence of the two frames has been demonstrated for some important 
observables \cite{Makino:1991sg,Deruelle:2010ht,Li:2014qwa,Gong:2011qe},
however, that may not be the whole story for a number of reasons:
\begin{itemize}
\item{Singularities (typical for super-acceleration) can lead to a 
breakdown of the mathematical equivalence between the two frames 
\cite{Briscese:2006xu,Bahamonde:2016wmz,Kamenshchik:2016gcy};}
\item{The non-gravitational sector of the theory knows the difference
because matter is minimally coupled in the Jordan frame whereas the coupling 
is highly non-minimal in the Einstein frame \cite{Domenech:2016yxd,
Kuusk:2016rso}; and}
\item{It can happen that the universe accelerates in one frame while 
decelerating in the other \cite{Capozziello:2006dj}.}
\end{itemize}
Nevertheless, it is expected that, for regular geometries, and in the absence 
of matter, the two frames are indeed equivalent. Studies of $f(R)$ inflation 
have been made in the Jordan frame \cite{Sebastiani:2013eqa,Artymowski:2015pna,
Odintsov:2015tka,Bamba:2014wda}, but the normal, and much easier approach, is 
to work in the Einstein frame.

Although we shall have to discuss the issue of frame dependence somewhat,
the purpose of this paper is to extend to $f(R)$ inflation a new formalism
for computing the scalar and tensor power spectra. The formalism is based
on first replacing the usual linear evolution equations for the mode functions
with nonlinear evolution equations for the norm-squared mode functions which
go into the power spectra \cite{Romania:2012tb}. This avoids the wasted effort
of keeping track of the irrelevant phase. We then factor out the exact solutions
which exist for constant first slow roll parameter, and derive a Green's
function solution for the residual factor \cite{Brooker:2015iya,Brooker:2016xkx}
which can be written for an {\it arbitrary inflationary geometry}. The power
inherent is this analytic functional representation has been recently exploited
to derive an improved version \cite{Brooker:2016imi} of the famous single scalar 
consistency relation \cite{Polarski:1995zn,GarciaBellido:1995fz,Sasaki:1995aw}.

In section 2 we show how primordial perturbations appear in the Jordan and
Einstein frames. Section 3 is devoted to the issue of using the power
spectra (when the tensor power spectrum is eventually resolved) to 
reconstruct either a scalar potential model or an $f(R)$ model which would
generate them. In section 4 we apply the new technique to two models of
$f(R)$ inflation. Our conclusions comprise section 5.

\section{Numerical Equality but Form Dependence}

The purpose of this section is to show that the scalar and tensor
perturbation fields of the Jordan and Einstein frames agree, but 
their power spectra nonetheless take highly different forms when 
expressed in terms of the geometrical quantities of each frame. We 
begin with a careful definition of the two frames, their backgrounds,
their natural gauges and their perturbation fields. We then give the
relation between the backgrounds and perturbations of each frame.
The Starobinsky model provides a nice illustration of frame dependence 
because the standard slow roll approximations for the power spectra are
valid in the Einstein frame but completely incorrect in the Jordan frame. 

\subsection{The Model in the Jordan Frame}

The (spacelike) metric of the Jordan frame is $g_{\mu\nu}$, which couples
minimally to matter and gives physical distances and times. The 
Lagrangian of this frame is,
\begin{equation}
\mathcal{L} = \frac{f(R) \sqrt{-g}}{16 \pi G} \; . \label{LJ}
\end{equation}
Its equation of motion is,
\begin{equation}
f'(R) R_{\mu\nu} - \frac12 f(R) g_{\mu\nu} + \Bigl[g_{\mu\nu} \square - 
D_{\mu} D_{\nu} \Bigr] f'(R) = 0 \; , \label{EOMJ}
\end{equation}
where $D_{\mu}$ represents the covariant derivative operator and $\square
\equiv g^{\mu\nu} D_{\mu} D_{\nu}$ is the covariant d'Alembertian.

The background geometry of the Jordan frame takes the form,
\begin{equation}
ds^2 = -dt^2 + a^2(t) d\vec{x} \!\cdot\! d\vec{x} = a^2(t) \Bigl[-d\eta^2
+ d\vec{x} \!\cdot\! d\vec{x}\Bigr] \; .
\end{equation}
One can see from (\ref{EOMJ}) one can see that this background obeys the 
equations,
\begin{eqnarray}
0 & \!\!\!=\!\!\! & -3(\dot{H} \!+\ H^2) f'\Bigl(R_0(t)\Bigr) + \frac12 
f\Bigl(R_0(t)\Bigr) + 3 H \partial_t f'\Bigl(R_0(t)\Bigr) \; , \label{E1} \\
0 & \!\!\!=\!\!\! & (\dot{H} \!+\! 3 H^2) f'\Bigl( R_0(t)\Bigr) - 
\frac12 f\Bigl( R_0(t)\Bigr) - (\partial_t^2 \!+\! 2 H \partial_t) 
f'\Bigl( R_0(t) \Bigr) \; . \quad \label{E2}
\end{eqnarray}
Here and henceforth $H(t) \equiv \dot{a}/a$ is the Hubble parameter of the
Jordan frame and $R_0(t) \equiv 6 \dot{H}(t) + 12 H^2(t)$ is the background 
value of the Ricci scalar. Adding (\ref{E1}) to (\ref{E2}) gives a relation
we shall exploit later,
\begin{equation}
\partial_t \Bigl[ f''(R_0) \dot{R}_0\Bigr] = -2 \dot{H} f'(R_0) + H f''(R_0)
\dot{R}_0 \; . \label{E3}
\end{equation}

The natural temporal gauge condition for the Jordan frame is $R(t,\vec{x}) = 
R_0(t)$ \cite{Jaime:2010kn}. In this gauge the $g_{00}$ and $g_{0i}$ 
components of the metric are constrained fields. The $g_{ij}$ components 
take the form,
\begin{equation}
g_{ij}(t,\vec{x}) = a^2(t) \times e^{2 \zeta(t,\vec{x})} \times
\Bigl[ e^{h(t,\vec{x})} \Bigr]_{ij} \qquad , \qquad h_{ii}(t,\vec{x}) = 0 
\; . \label{Jframegij}
\end{equation}
Note that requiring $h_{ii} = 0$ is not a gauge condition but rather how one
defines the breakup between $\zeta$ and $h_{ij}$. The spatial gauge condition 
is,
\begin{equation}
\partial_i h_{ij}(t,\vec{x}) = 0 \; .
\end{equation} 

The homogeneity and isotropy of the Jordan frame background implies that
the perturbation fields have the following free field expansions,
\begin{eqnarray}
h_{ij}(t,\vec{x}) & = & \sqrt{32 \pi G} \int \!\! \frac{d^3k}{(2\pi)^3} 
\sum_{\lambda = \pm} \Biggl\{ u(t,k) e^{i \vec{k} \cdot \vec{x}} 
\epsilon_{ij}(\vec{k},\lambda) \alpha(\vec{k},\lambda) + {\rm c.c.} 
\Biggr\} \; , \qquad \\
\zeta(t,\vec{x}) & = & \sqrt{4\pi G} \int \!\! \frac{d^3k}{(2\pi)^3} 
\Biggl\{ v(t,k) e^{i \vec{k} \cdot \vec{x}} \beta(\vec{k}) + {\rm c.c.} 
\Biggr\} \; .
\end{eqnarray}
The polarization tensor $\epsilon_{ij}(\vec{k},\lambda)$ obeys the same 
relations as in flat space, and is identical to the flat space result,
\begin{equation}
k_i \epsilon_{ij} = 0 = \epsilon_{ii} \qquad , \qquad 
\epsilon_{ij}(\vec{k},\kappa) \epsilon^*_{ij}(\vec{k},\lambda) =
\delta_{\kappa \lambda} \; .
\end{equation}
The creation and annihilation operators also obey the flat space relations,
\begin{equation}
\Bigl[ \alpha(\vec{k},\kappa) , \alpha^{\dagger}(\vec{p},\lambda)\Bigr] =
\delta_{\kappa \lambda} (2\pi)^3 \delta^3( \vec{k} \!-\! \vec{p}) \quad ,
\quad \Bigl[ \beta(\vec{k}) , \beta^{\dagger}(\vec{p}) \Bigr] = 
(2\pi)^3 \delta^3(\vec{k} \!-\! \vec{p}) \; .
\end{equation}
It is best to define the (tree order) power spectra as the asymptotic late 
time forms of equal-time correlators,
\begin{eqnarray}
\Delta^2_h(t,k) & \!\!\!\!\equiv\!\!\!\! & \frac{k^3}{2\pi^2} \! \int \!\! 
d^3x \, e^{-i \vec{k} \cdot \vec{x}} \Bigl\langle \Omega \Bigl\vert 
h_{ij}(t,\vec{x}) h_{ij}(t,\vec{0}) \Bigr\vert \Omega \Bigr\rangle 
\nonumber \\
& & \hspace{5cm} = \frac{k^3}{2 \pi^2} \!\times\! 32\pi G \!\times\! 2 
\!\times\! \vert u(t,k)\vert^2 \; , \label{powerh} \\
\Delta^2_{\mathcal{R}}(t,k) & \!\!\!\!\equiv\!\!\!\! & \frac{k^3}{2\pi^2} 
\! \int \!\! d^3x \, e^{-i \vec{k} \cdot \vec{x}} \Bigl\langle \Omega 
\Bigl\vert \zeta(t,\vec{x}) \zeta(t,\vec{0}) \Bigr\vert \Omega 
\Bigr\rangle = \frac{k^3}{2 \pi^2} \!\times\! 4\pi G \!\times\! \vert 
v(t,k)\vert^2 \; , \label{powerR} \qquad
\end{eqnarray}

The equations obeyed by the tensor mode function $u(t,k)$ are fairly easy to 
read off by linearizing (\ref{EOMJ}) and applying canonical quantization,
\begin{equation}
\ddot{u} + \Bigl( 3 H \!+\! \frac{f''(R_0) \dot{R}_0}{f'(R_0)}\Bigr) \dot{u} 
+ \frac{k^2}{a^2} \, u = 0 \qquad , \qquad u \dot{u}^* - \dot{u} u^* = 
\frac{i}{f'(R_0) a^3} \; . \label{ueqn}
\end{equation}
Obtaining the scalar mode equations is much more difficult because one must
first solve the constraints. A long calculation reveals that $v(t,k)$ obeys,
\begin{equation}
\ddot{v} + \Bigl( 3 H \!+\! \frac{f''(R_0) \dot{R}_0}{f'(R_0)} \!+\! 
\frac{\dot{E}}{E} \Bigr) \dot{v} + \frac{k^2}{a^2} \, v = 0 \quad , 
\quad v \dot{v}^* - \dot{v} v^* = \frac{i}{E f'(R_0) a^3} \; , \label{veqn}
\end{equation}
where the function $E(t)$ is,
\begin{equation}
E = \frac{ 3 (\frac{f''(R_0) \dot{R}_0}{2 f'(R_0) H})^2}{ (1 \!+\! 
\frac{f''(R_0) \dot{R}_0}{2 f'(R_0) H})^2} \; .
\end{equation}
Differential equations such as (\ref{ueqn}-\ref{veqn}) define the mode 
functions up to initial conditions. The usual (Bunch-Davies-like) initial
conditions are that the WKB forms apply in the distant past,
\begin{eqnarray}
u(t,k) & \longrightarrow & \frac1{\sqrt{2 k f'(R_0(t)) a^2(t)}} \, 
\exp\Bigl[ -i k \! \int_{t_i}^{t}\! \frac{dt'}{a(t')} \Bigr] \; , 
\label{initialu} \\
v(t,k) & \longrightarrow & \frac1{\sqrt{2 k E(t) f'(R_0(t)) a^2(t)}} \, 
\exp\Bigl[ -i k \! \int_{t_i}^{t}\! \frac{dt'}{a(t')} \Bigr] \; .
\label{initialv}
\end{eqnarray}
One can see from (\ref{ueqn}-\ref{veqn}) that the mode functions must
approach constants when the term $k^2/a^2(t)$ becomes insignificant. 
Those constants can be found by using (\ref{ueqn}-\ref{veqn}) to evolve
$u(t,k)$ and $v(t,k)$ from their initial forms 
(\ref{initialu}-\ref{initialv}). Substituting those constants into the 
time-dependent power spectra (\ref{powerh}-\ref{powerR}) gives the
model's predictions for the primordial power spectra. 

\subsection{The Model in the Einstein Frame}

The transformation from the Jordan frame to the Einstein frame is 
effected by first introducing an auxiliary scalar $\phi$ which obeys 
the equation,
\begin{equation}
\phi = f'(R) \qquad \Longleftrightarrow \qquad R = \mathcal{R}(\phi) \; .
\label{step1}
\end{equation}
We then construct a potential $U(\phi)$ by Legendre transforming,
\begin{equation}
U(\phi) \equiv \phi \mathcal{R}(\phi) - f\Bigl( \mathcal{R}(\phi)\Bigr)
\qquad \Longleftrightarrow \qquad U'(\phi) = \mathcal{R}(\phi) \; .
\label{step2}
\end{equation} 
The Einstein frame Lagrangian is,
\begin{equation}
\widetilde{\mathcal{L}} = \frac{1}{16 \pi G} \Bigl[ \phi R - 
U(\phi)\Bigr] \sqrt{-g} \; . \label{LE}
\end{equation}
The two field equations associated with (\ref{LE}) are,
\begin{eqnarray}
0 & = & R - U'(\phi) \; , \label{EOME1} \\
0 & = & \phi R_{\mu\nu} -\frac12 \Bigl[ \phi R \!-\! U(\phi)\Bigr] +
\Bigl[ g_{\mu\nu} \square \!-\! D_{\mu} D_{\nu} \Bigr] \phi \; .
\label{EOME2}
\end{eqnarray}
Of course (\ref{EOME1}) reproduces (\ref{step1}), whereupon we recognize
(\ref{EOME2}) as the Jordan frame equation (\ref{EOMJ}).

We reach the final form of the Einstein frame by making a field 
redefinition which is the conformal transformation,
\begin{eqnarray}
\widetilde{g}_{\mu\nu} \equiv \phi g_{\mu\nu} \qquad & \Longleftrightarrow 
& \qquad g_{\mu\nu} = \exp\Biggl[-\sqrt{\frac{16\pi G}{3}} \, \varphi\Biggr]
\widetilde{g}_{\mu\nu} \; , \label{step3a} \\
\varphi \equiv \sqrt{\frac{3}{16\pi G}} \, \ln(\phi) \qquad & 
\Longleftrightarrow & \qquad \phi = \exp\Biggl[ \sqrt{\frac{16\pi G}{3}} 
\, \varphi\Biggr] \; . \label{step3b}
\end{eqnarray}
Substituting (\ref{step3a}-\ref{step3b}) in (\ref{LE}) gives the classic 
form of a minimally coupled scalar,
\begin{equation}
\widetilde{\mathcal{L}} = \frac{\widetilde{R} \sqrt{-\widetilde{g}}}{16\pi G} 
-\frac12 \partial_{\mu} \varphi \partial_{\nu} \varphi \widetilde{g}^{\mu\nu}
\sqrt{-\widetilde{g}} - V(\varphi) \sqrt{-\widetilde{g}} \; . \label{LEa}
\end{equation}
where the scalar potential is,
\begin{equation}
V(\varphi) \equiv \frac1{16 \pi G} \, \exp\Biggl[ -2\sqrt{\frac{16 \pi G}{3}}
\, \varphi\Biggr] U\Biggl( \exp\Biggl[ \sqrt{\frac{16 \pi G}{3}} \, \varphi
\Biggr] \Biggr) \; . \label{LEb}
\end{equation}

The background geometry of the Einstein frame takes the form,
\begin{equation}
d\widetilde{s}^2 = -d\widetilde{t}^2 + \widetilde{a}^2(\widetilde{t}) 
d\vec{x} \!\cdot\! d\vec{x} = \widetilde{a}^2(\widetilde{t}) \Bigl[-d\eta^2
+ d\vec{x} \!\cdot\! d\vec{x}\Bigr] \; . \label{ds2E}
\end{equation}
It relates to the background scalar field $\varphi_0(\widetilde{t})$ through
the Einstein equations,
\begin{eqnarray}
3 \widetilde{H}^2(\widetilde{t}) & = & 8\pi G \Bigl[\frac12 
\dot{\varphi}_0^2(\widetilde{t}) + V\Bigl( \varphi_0(\widetilde{t})\Bigr)
\Bigr] \; , \\
-2 \dot{\widetilde{H}}(\widetilde{t}) -3 \widetilde{H}^2(\widetilde{t}) 
& = & 8\pi G \Bigl[\frac12 \dot{\varphi}_0^2(\widetilde{t}) - 
V\Bigl( \varphi_0(\widetilde{t})\Bigr) \Bigr] \; .
\end{eqnarray}

The natural temporal gauge condition in the Einstein frame is 
$\varphi(\widetilde{t},\vec{x}) = \varphi_0(\widetilde{t})$
\cite{Salopek:1988qh}. In this gauge the $\widetilde{g}_{00}$ and 
$\widetilde{g}_{0i}$ components of the metric are constrained fields and 
the spatial components take the form,
\begin{equation}
\widetilde{g}_{ij}(\widetilde{t},\vec{x}) \equiv \widetilde{a}^2(\widetilde{t})
\times e^{2 \widetilde{\zeta}(\widetilde{t},\vec{x})} \times \Bigl[ 
e^{\widetilde{h}(\widetilde{t},\vec{x})} \Bigr]_{ij} \qquad , \qquad
\widetilde{h}_{ii} = 0 \; . \label{Eframegij}
\end{equation}
Note that requiring $\widetilde{h}_{ij}(\widetilde{t},\vec{x})$ to be 
traceless is not a gauge condition but rather part of the definition of
$\widetilde{\zeta}(\widetilde{t},\vec{x})$. The true spatial gauge 
condition is the transversality of $\widetilde{h}_{ij}(\widetilde{t},\vec{x})$,
\begin{equation}
\partial_i \widetilde{h}_{ij}(\widetilde{t},\vec{x}) = 0 \; .
\end{equation}

Homogeneity and isotropy are also symmetries in the Einstein frame so we
can expand the perturbation fields the same way as in the Jordan frame,
only with different mode functions,
\begin{eqnarray}
\widetilde{h}_{ij}(\widetilde{t},\vec{x}) & = & \sqrt{32 \pi G} \int \!\! 
\frac{d^3k}{(2\pi)^3} \sum_{\lambda = \pm} \Biggl\{ \widetilde{u}(
\widetilde{t},k) e^{i \vec{k} \cdot \vec{x}} \epsilon_{ij}(\vec{k},\lambda) 
\alpha(\vec{k},\lambda) + {\rm c.c.} \Biggr\} \; , \qquad \\
\widetilde{\zeta}(\widetilde{t},\vec{x}) & = & \sqrt{4 \pi G} \int \!\! 
\frac{d^3k}{(2\pi)^3} \Biggl\{ \widetilde{v}(\widetilde{t},k)
e^{i \vec{k} \cdot \vec{x}} \beta(\vec{k}) + {\rm c.c.} \Biggr\} \; .
\end{eqnarray}
Note that the polarization tensor of the Einstein frame is identical to 
that of the Jordan frame, as are the creation and annihilation operators. 
The time dependent power spectra are defined in the same way as for the
Jordan frame to give,
\begin{eqnarray}
\widetilde{\Delta}^2_{h}(\widetilde{t},k) & \equiv & \frac{k^3}{2 \pi^2}
\!\times\! 32\pi G \!\times\! 2 \!\times\! \vert \widetilde{u}(
\widetilde{t},k) \vert^2 \; , \label{powerhtilde} \\
\widetilde{\Delta}^2_{\mathcal{R}}(\widetilde{t},k) & \equiv & 
\frac{k^3}{2 \pi^2} \!\times\! 4\pi G \!\times\! \vert \widetilde{v}(
\widetilde{t},k) \vert^2 \; . \label{powerRtilde}
\end{eqnarray}

By solving the constraint equations and employing canonical quantization 
one finds that the mode functions obey the following equations and 
Wronskian normalization conditions,
\begin{eqnarray}
\Bigl[ \frac{\partial^2}{\partial \widetilde{t}^2} + 3 \widetilde{H}
\frac{\partial}{\partial \widetilde{t}} + \frac{k^2}{\widetilde{a}^2}
\Bigr] \widetilde{u} = 0 \qquad & , & \qquad \widetilde{u} 
\frac{\partial \widetilde{u}^*}{\partial \widetilde{t}} - 
\frac{\partial \widetilde{u}}{\partial \widetilde{t}} \, \widetilde{u}^* = 
\frac{i}{\widetilde{a}^3} \; , \label{utilde} \\
\Bigl[ \frac{\partial^2}{\partial \widetilde{t}^2} + \Bigl(3 \widetilde{H}
\!+\! \frac1{\widetilde{\epsilon}} \frac{d \widetilde{\epsilon}}{d 
\widetilde{t}} \Bigr) \frac{\partial}{\partial \widetilde{t}} + 
\frac{k^2}{\widetilde{a}^2} \Bigr] \widetilde{v} = 0 \qquad & , & \qquad
\widetilde{v} \frac{\partial \widetilde{v}^*}{\partial \widetilde{t}} - 
\frac{\partial \widetilde{v}}{\partial \widetilde{t}} \, \widetilde{v}^* = 
\frac{i}{\widetilde{\epsilon} \widetilde{a}^3} \; . \qquad \label{vtilde}
\end{eqnarray}
The assumption of Bunch-Davies-like vacuum corresponds to the following
asymptotic early time forms,
\begin{eqnarray}
\widetilde{u}(\widetilde{t},k) & \longrightarrow & \frac1{\sqrt{2 k 
\widetilde{a}^2(\widetilde{t})}} \, \exp\Bigl[ -i k \!
\int_{\widetilde{t}_i}^{\widetilde{t}}\! \frac{dt'}{\widetilde{a}(t')}
\Bigr] \; , \label{initialutilde} \\
\widetilde{v}(\widetilde{t},k) & \longrightarrow & \frac1{\sqrt{2 k 
\widetilde{\epsilon}(\widetilde{t}) \widetilde{a}^2(\widetilde{t})}} \, 
\exp\Bigl[ -i k \! \int_{\widetilde{t}_i}^{\widetilde{t}}\! \frac{dt'}{
\widetilde{a}(t')} \Bigr] \; . \label{initialvtilde}
\end{eqnarray}
The model's predictions for the primordial power spectra are obtained
by using (\ref{utilde}-\ref{vtilde}) to evolve $\widetilde{u}(
\widetilde{t},k)$ and $\widetilde{v}(\widetilde{t},k)$ from their
initial forms (\ref{initialutilde}-\ref{initialvtilde}) to find their 
late time constant values, and then substituting these constants into
the time dependent power spectra (\ref{powerhtilde})-\ref{powerRtilde}).

\subsection{Relating Backgrounds and Perturbation Fields}

Comparison of expression (\ref{Jframegij}) with (\ref{Eframegij}), and 
relations (\ref{step3a}-\ref{step3b}), implies that the perturbation 
fields agree between the two frames \cite{Gong:2011qe},
\begin{eqnarray}
\zeta(t,\vec{x}) & = & \widetilde{\zeta}(\widetilde{t},\vec{x}) \; , 
\label{ztoztilde} \\
h_{ij}(t,\vec{x}) & = & \widetilde{h}_{ij}(\widetilde{t},\vec{x}) \; 
\label{htohtilde} .
\end{eqnarray}
This means that the scalar and tensor power spectra also agree 
numerically between the two frames. However, expressions for those 
power spectra are quite frame dependent because the expansion 
histories and co-moving times of the two frames do not agree,
\begin{eqnarray}
a(t) = \exp\Bigl[- \sqrt{\frac{4 \pi G}{3}} \, \varphi(\widetilde{t}) 
\Bigr] \!\times\! \widetilde{a}(\widetilde{t}) \;\; & 
\Longleftrightarrow & \;\; \widetilde{a}(\widetilde{t}) = 
\sqrt{f'\Bigl( R_0(t)\Bigr)} \!\times\! a(t) \; , \quad 
\label{atoatilde} \\
dt = \exp\Bigl[- \sqrt{\frac{4 \pi G}{3}} \, \varphi(\widetilde{t}) 
\Bigr] \!\times\! d\widetilde{t} \;\; & \Longleftrightarrow & \;\; 
d\widetilde{t} = \sqrt{f'\Bigl( R_0(t)\Bigr)} \!\times\! dt \; . 
\quad \label{dttodttilde}
\end{eqnarray}
It follows that the Hubble parameter of the Einstein frame is,
\begin{eqnarray}
\widetilde{H}(\widetilde{t}) & \equiv & \frac{d}{d \widetilde{t}} 
\ln\Bigl[ \widetilde{a}(\widetilde{t})\Bigr] = 
\frac1{\sqrt{f'(R_0(t))}} \frac{d}{d t} \ln\Bigl[ \sqrt{
f'(R_0(t))} \, a(t)\Bigr] \; , \qquad \\
& = & \frac{H}{\sqrt{f'(R_0)}} \Biggl[1 + \frac{ f''(R_0) \dot{R}_0}{
2 f'(R_0) H} \Biggr] \; . \label{Htilde}
\end{eqnarray}
Using relation (\ref{E3}) the first slow roll parameter is,
\begin{equation}
\widetilde{\epsilon}(\widetilde{t}) \equiv \frac{d}{d\widetilde{t}}
\frac1{\widetilde{H}(\widetilde{t})} =
\frac1{\sqrt{f'(R_0)}} \frac{d}{d t} \Biggl[ \frac{\sqrt{f'(R_0)}}{
H + \frac{f''(R_0) \dot{R}_0}{2 f'(R_0)} }\Biggr] 
= \frac{3 (\frac{f''(R_0) \dot{R}_0}{2 f'(R_0) H})^2}{[1 + 
\frac{f''(R_0) \dot{R}_0}{2 f'(R_0) H}]^2} \; . \label{epstilde}
\end{equation}
Both parameters depend critically on the function $X$,
\begin{equation}
X \equiv \frac{f''(R_0) \dot{R}_0}{2 f'(R_0) H} = 
-\frac{f''(R_0) R_0}{f'(R_0)} \Bigl[ \epsilon + 
\frac{\dot{\epsilon}}{2 (2 \!-\! \epsilon) H}\Bigr] \; , \label{Xdef}
\end{equation}
where the final form on the right follows from $R_0 = 6 (2-\epsilon) 
H^2$ and hence,
\begin{equation} 
\dot{R}_0 = -12 \epsilon (2 \!-\! \epsilon) H^3 - 6 \dot{\epsilon} H^2
= -2 H (\epsilon R_0 \!+\! 3 \dot{\epsilon} H) \; .
\end{equation}

Combining relations (\ref{Htilde}-\ref{epstilde}) with the usual slow roll 
results for the power spectra in the Einstein frame (and hence also in the
Jordan frame) gives,
\begin{eqnarray}
\Delta^2_{\mathcal{R}}(k) \simeq \frac{G \widetilde{H}^2}{\pi \widetilde{
\epsilon}} & = & \frac{G H^2 (1 \!+\! X)^4}{3\pi f'(R_0) X^2} \; , 
\label{DR} \\
\Delta^2_{h}(k) \simeq \frac{16}{\pi} G \widetilde{H}^2 & = &
\frac{16 G H^2 (1 \!+\! X)^2}{\pi f'(R_0)} \; . \label{Dh}
\end{eqnarray}
Therefore, the tensor-to-scalar ratio is,
\begin{equation}
r(k) \equiv \frac{\Delta^2_{h}(k)}{\Delta^2_{\mathcal{R}}(k)} \approx
16 \widetilde{\epsilon} = \frac{48 X^2}{(1 \!+\! X)^2} \; . \label{r}
\end{equation}
Successful models of $f(R)$ inflation typically have $f''(R_0) R_0/f'(R_0) 
\sim 1$, so relation (\ref{Xdef}) implies $X \sim -\epsilon$. Substituting
into relation (\ref{r}) means that slow roll inflation in the Einstein
frame, with $r \approx 16 \widetilde{\epsilon}$, typically implies $r 
\approx 48 \epsilon^2$ when expressed using the Jordan frame geometry.

\subsection{Starobinsky Inflation}

Starobinsky inflation corresponds to,
\begin{equation}
f(R) = R + \frac{8 \pi G R^2}{6 M^2} \quad \Longrightarrow \quad 
f'(R) = 1 + \frac{16 \pi G R}{6 M^2} \quad \Longrightarrow \quad
f''(R) = \frac{16 \pi G}{6 M^2} \; . \label{AAS}
\end{equation}
Substituting (\ref{AAS}) into the background equations (\ref{E1}-\ref{E2})
reveals a good approximate solution with,
\begin{equation}
\dot{H}(t) \simeq -\frac{M^2}{48 \pi G} \equiv -\epsilon_i H_i^2 \; ,
\end{equation}
where $H_i$ and $\epsilon_i$ are the initial values of the Hubble and
first slow roll parameters. Hence the various geometrical parameters are,
\begin{eqnarray}
\epsilon(t) & \simeq & \frac{\epsilon_i}{[1 \!-\! \epsilon_i H_i 
\Delta t]^2} \; , \\
H(t) & \simeq & H_i [1 \!-\! \epsilon_i H_i \Delta t] \; , \\
a(t) & \simeq & a_i \exp\Bigl[ H_i \Delta t \!-\! \frac12 \epsilon_i 
(H_i \Delta t)^2\Bigr] \; .
\end{eqnarray}
Expressing these parameters in terms of the number of e-foldings $n$ 
from the start of inflation gives,
\begin{equation}
\epsilon = \frac{\epsilon_i}{1 \!-\! 2 \epsilon_i n} \quad , \quad
H = H_i \sqrt{1 \!-\! 2 \epsilon_i n} \quad , \quad a = a_i e^{n} \; .
\end{equation}

\begin{figure}[ht]
\includegraphics[width=10cm,height=8cm]{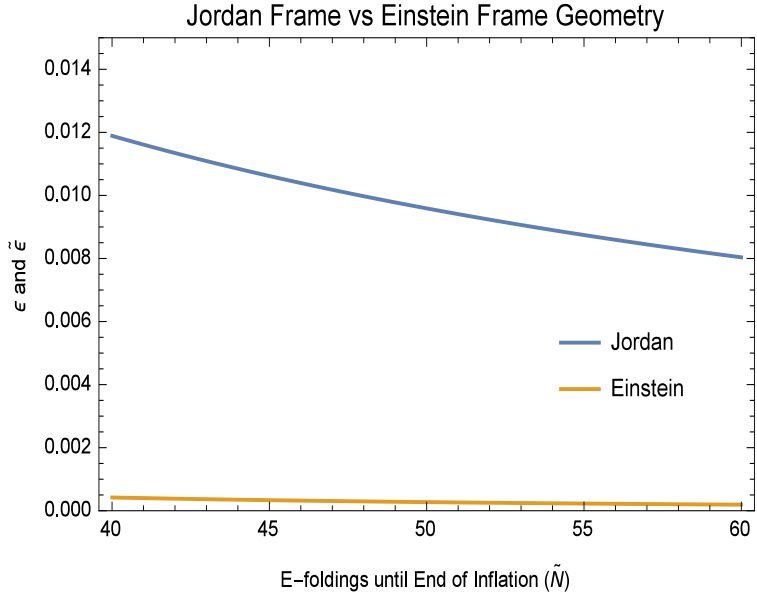}
\caption{Comparison of first slow roll parameter in the two frames for
Starobinsky inflation. The blue curve gives Jordan frame result 
$\epsilon$ whereas the yellow curve shows the much smaller Einstein 
frame result $\widetilde{\epsilon}$ of expression (\ref{epstilde}).}
\label{1stslowroll}
\end{figure}

Under the usual assumption that $0 < \epsilon_i \ll 1$ we have,
\begin{equation}
f''\Bigl(R_0(t)\Bigr) R_0(t) \simeq \frac{2}{3 \epsilon(t)} \simeq
f'\Bigl( R_0(t)\Bigr) \; .
\end{equation}
Substituting into relation(\ref{Xdef}) implies,
\begin{equation}
X(t) \simeq -\epsilon(t) \; .
\end{equation}
Hence the first slow roll parameter of the Einstein frame (\ref{epstilde})
is much smaller than the first slow roll parameter of the Jordan frame, as
depicted in Fig.~\ref{1stslowroll}. The power spectra and their ratio are,
\begin{eqnarray}
\Delta^2_{\mathcal{R}}(k) & \simeq & \frac{G H^2}{2 \pi \epsilon} 
\simeq \frac{G H_i^2}{2\pi \epsilon_i} \Bigl[1 \!-\! 2 \epsilon_i n_k
\Bigr]^2 \; , \label{R2DR} \\
\Delta^2_{h}(k) & \simeq & \frac{24}{\pi} \, G H^2 \epsilon \simeq 
\frac{24}{\pi} \, G H_i^2 \epsilon_i \; , \label{R2Dh} \\
r(k) & \simeq & 48 \epsilon^2 \simeq \frac{48 \epsilon_i^2}{(1 \!-\!
2\epsilon_i n_k)^2} \; , \label{R2r}
\end{eqnarray}
where $n_k \simeq \ln(k/a_i H_i)$ is the e-folding of first horizon 
crossing. This model actually obeys the famous single-scalar consistency
relation \cite{Polarski:1995zn,GarciaBellido:1995fz,Sasaki:1995aw} but 
one would need to carry the expansion of $\Delta^2_{h}(k) \simeq 
\frac{24}{\pi} G H^2 \epsilon \times (1 - 3 \epsilon + \dots)$ one more
order to give a nonzero result for the tensor spectral index. However,
relations (\ref{R2DR}-\ref{R2r}) deviate extensively from the usual slow 
roll results when expressed in terms of the Jordan frame geometry.

Starobinsky inflation obeys the general rule of $f(R)$ inflation that
its power spectra are numerically the same, for fixed wave number $k$, in 
both Jordan and Einstein frames. However, what this ``$k$'' means 
geometrically is very different in the two frames. One way to see the
difference is by expressing the spectra in terms of the number of 
e-foldings until the end of inflation. From relation (\ref{atoatilde}) we 
infer,
\begin{equation}
\widetilde{a}(\widetilde{t}) \equiv \widetilde{a}_i e^{\widetilde{n}} 
\simeq \frac{a(t)}{\sqrt{\frac32 \epsilon(t)}} \quad \Longrightarrow \quad
\widetilde{a}_i \simeq \frac{a_i}{\sqrt{ \frac32 \epsilon_i}} \; , \;
\widetilde{n} \simeq n + \frac12 \ln(1 \!-\! 2 \epsilon_i n) \; .
\end{equation}
Inflation ends at $n_{\rm end} \simeq \frac1{2 \epsilon_i} - \frac12$,
which corresponds to $\widetilde{n}_{\rm end} \simeq \frac1{2 \epsilon_i}
+ \frac12 \ln(\epsilon_i)$. The number of Jordan frame e-foldings until
the end of inflation is $N \equiv n_{\rm end} - n$, so the number of
Einstein frame e-foldings until the end of inflation is,
\begin{equation}
\widetilde{N} \equiv \widetilde{n}_{\rm end} - \widetilde{n} \simeq
N - \frac12 \ln(1 \!+\! 2 N) \; . \label{NtoNtilde}
\end{equation}
Therefore, a feature which occurs $N = 50$ Jordan frame e-foldings before 
the end of inflation appears at about $\widetilde{N} \simeq 47.7$ e-foldings 
Einstein frame e-foldings before the end of inflation.

\section{Constructing Models from Power Spectra}

Because the perturbation fields of the Einstein and Jordan frames are
identical, the power spectra in each frame are the same functions of 
the wave number $k$. Given only these functions $\Delta^2_{\mathcal{R}}(k)$
and $\Delta^2_{h}(k)$, one cannot tell whether primordial inflation was 
driven by a scalar potential model or by an $f(R)$ model. The purpose of 
this section is to explain how to reconstruct either sort of model. We 
begin by using $\Delta^2_{\mathcal{R}}(k)$ and $\Delta^2_{h}(k)$ to infer 
the scalar potential model which would produce them. We then construct the 
$f(R)$ model that would produce the same results.

\subsection{Reconstructing a Scalar Potential Model}

If the inflationary expansion history $a(t)$ is driven by the potential of a 
single, minimally coupled scalar then the resulting (tree order) scalar and 
tensor power spectra can be expressed in terms of the geometry near the 
time $t_k$ of first crossing, $k \equiv H(t_k) a(t_k)$. The exact formulae
take the form of leading slow roll results, times local slow roll corrections, 
multiplied by nonlocal factors \cite{Brooker:2015iya,Brooker:2016xkx},
\begin{eqnarray}
\Delta^2_{\mathcal{R}}(k) & = & \frac{G H^2(t_k)}{\pi \epsilon(t_k)} 
\!\times\! C\Bigl( \epsilon(t_k)\Bigr) \!\times\! \mathcal{S}(k) \; , 
\label{fullDR} \\
\Delta^2_{h}(k) & = & \frac{16}{\pi} \, G H^2(t_k) \!\times\! C\Bigl( 
\epsilon(t_k)\Bigr) \!\times\! \mathcal{C}(k) \; . \label{fullDh}
\end{eqnarray}
The local slow roll correction $C(\epsilon)$ is a monotonically deceasing 
function well approximated by $1-\epsilon$ (see Figure~2 of 
\cite{Brooker:2015iya}),
\begin{equation}
C(\epsilon) \equiv \frac1{\pi} \Gamma^2\Bigl( \frac12 \!+\! 
\frac1{1 \!-\! \epsilon}\Bigr) \Bigl[ 2 (1 \!-\! \epsilon)
\Bigr]^{\frac2{1-\epsilon}} \approx 1 - \epsilon \; . \label{Cdef}
\end{equation}
The nonlocal correction factors, $\mathcal{S}(k)$ and $\mathcal{C}(k)$,
are unity for $\dot{\epsilon} = 0$ and depend in a completely known way
\cite{Brooker:2015iya,Brooker:2016xkx} upon conditions only a few 
e-foldings before and after $t_k$.

It would be simple enough to give an successive approximation technique
for exactly reconstructing $H^2(t_k)$ from the full expressions 
(\ref{fullDR}-\ref{fullDh}) but we will here work with just the leading 
slow roll results. First, express $\Delta^2_{\mathcal{R}}(k)$ as a 
differential equation for $H(t_k)$,
\begin{equation}
\Delta^2_{\mathcal{R}}(k) \simeq \frac{G H^2(t_k)}{\pi \epsilon(t_k)}
\quad \Longrightarrow \quad \frac1{H(t_k)} \, \frac{d}{d t_k} \,
\frac1{H^2(t_k)} \simeq \frac{2G}{\pi} \frac1{\Delta^2_{\mathcal{R}}(k)} 
\; . \label{stepA}
\end{equation}
Now multiply by $H(t_k) dt_k \simeq dk/k$, integrate to solve for
$H^2(t_k)$, and express the integration constant in terms of the 
leading slow roll result for $\Delta^2_{h}(k)$,
\begin{equation}
H^2(t_k) \simeq \frac{H^2(t_*)}{1 \!+\! \frac{2 G H^2(t_*)}{\pi} 
\int_{k_*}^{k} \!\! \frac{dk'}{k'} \frac1{\Delta^2_{\mathcal{R}}(k')}}
\simeq \frac{\frac{\pi}{16 G} \, \Delta^2_{h}(k_*)}{1 + \frac18 r(k_*)
\int_{k_*}^{k} \frac{dk'}{k'} \frac{\Delta^2_{\mathcal{R}}(k_*)}{
\Delta^2_{\mathcal{R}}(k')}} \; .  \label{stepB}
\end{equation} 
One finds the scale factor by,
\begin{equation}
a(t_k) = \frac{k}{H(t_k)} \; . \label{recona}
\end{equation}
The construction is completed by integrating the differential relation
$H(t_k) dt_k \simeq dk/k$ and then inverting to solve for $k(t)$,
\begin{equation}
t = t_* + \int_{k_*}^{k} \!\! \frac{dk'}{k' H(t_{k'})} \qquad 
\Longleftrightarrow k=k(t) \; . \label{stepC}
\end{equation}

Of course these operations would have to be performed numerically, 
but we stress that, by going beyond the leading slow roll forms, the 
reconstruction could be accomplished to a precision limited only by 
the quality of the data for $\Delta^2_{\mathcal{R}}(k)$ and 
$\Delta^2_{h}(k)$. Note also that the construction depends much
more heavily on the well-measured scalar power spectrum, with its
tensor cousin used only to supply integration constants. By comparing
this reconstruction with $\Delta^2_{h}(k)$, when it is finally resolved, 
one can test the consistency of assuming single scalar inflation
\cite{Brooker:2016imi}.

Given the expansion history $a(t)$ and its derivatives, we can apply
a well known construction \cite{Tsamis:1997rk,Saini:1999ba,Nojiri:2005pu,
Capozziello:2005mj,Woodard:2006nt,Guo:2006ab} to find the scalar and its 
potential from the two nontrivial Einstein equations,
\begin{eqnarray}
3 H^2(t) & = & 8 \pi G \Bigl[ \frac12 \dot{\varphi}^2(t) + V\Bigl( 
\varphi(t)\Bigr) \Bigr] \; , \label{reconE1} \\
-2\dot{H}(t) -3 H^2(t) & = & 8 \pi G \Bigl[ \frac12 \dot{\varphi}^2(t) - 
V\Bigl( \varphi(t)\Bigr) \Bigr] \; , \label{reconE2}
\end{eqnarray}
By adding (\ref{reconE1}) to (\ref{reconE2}) we can reconstruct the scalar, 
up to its initial value and an arbitrary sign choice,
\begin{equation}
-2\dot{H}(t) = 8\pi G \dot{\varphi}^2(t) \quad \Longrightarrow \quad 
\varphi(t) = \varphi(t_i) \pm \int_{t_i}^{t} \!\! ds \, \sqrt{ \frac{-2
H(s)}{8\pi G}} \; . \label{reconphi}
\end{equation}
Expression (\ref{reconphi}) makes sense as long as $\dot{H}(t) < 0$, which
is the usual case. Under the same assumption, the scalar $\varphi(t)$ is a 
monotonically growing or falling function of time, and we can invert 
(\ref{reconphi}) to find $t(\varphi)$. The final step is substituting this 
expression into the difference of (\ref{reconE1}) and (\ref{reconE2}) in 
order to reconstruct the potential,
\begin{equation}
V(\varphi) = \frac{\dot{H}(t(\varphi)) + 3 H^2(t(\varphi))}{16 \pi G} 
\; . \label{reconV}
\end{equation}

\subsection{Reconstructing an $f(R)$ Model}

The previous subsection explained how the power spectra could be used to 
reconstruct a scalar potential model which would produce the observed
power spectra $\Delta^2_{\mathcal{R}}(k)$ and $\Delta^2_{h}(k)$. Suppose
that this has been done this. To find the $f(R)$ model which would produce
the very same power spectra, one begins by regarding the reconstructed 
expansion history (\ref{recona}) as the Einstein frame scale factor 
$\widetilde{a}(\widetilde{t})$ of some $f(R)$ model, expressed as a 
function of the Einstein frame time $\widetilde{t}$. Similarly, consider
the reconstructed scalar (\ref{reconphi}) as the Einstein frame scalar 
$\varphi(\widetilde{t})$, also expressed as a function of $\widetilde{t}$. 

The next step is to reconstruct the geometry of the Jordan frame. This
is accomplished by integrating equation (\ref{dttodttilde}) and inverting 
to express the Einstein frame time as a function of the Jordan frame time,
\begin{equation}
t = t_{i} + \int_{\widetilde{t}_i}^{\widetilde{t}} \!\! ds \,
\exp\Biggl[ -\sqrt{\frac{4 \pi G}{3}} \, \varphi(s)\Biggr] \quad 
\Longrightarrow \quad \widetilde{t}(t) \; .
\end{equation}
Now substitute into relation (\ref{atoatilde}) to find the Jordan frame
expansion history,
\begin{equation}
a(t) = \exp\Biggl[ -\sqrt{\frac{4 \pi G}{3}} \, \varphi\Bigl(
\widetilde{t}(t)\Bigr) \Biggr] \!\times\! \widetilde{a}\Bigl( 
\widetilde{t}(t)\Bigr) \; . 
\end{equation}
Of course this gives us the Hubble parameter $H(t)$ and the first slow
roll parameter $\epsilon(t)$ as well.

The final step is to reconstruct the function $f(R)$. First, invert the
relation for $R_0(t)$ to express time as a function of the Ricci scalar,
\begin{equation}
R_0(t) = 6 \Bigl[ 2 \!-\! \epsilon(t)\Bigr] H^2(t) \qquad 
\Longleftrightarrow \qquad t(R) \; . \label{tofR}
\end{equation}
Now note that the differential of the Ricci scalar is,
\begin{equation}
dR_0(t) = \Biggl\{ -12 \epsilon(t) \Bigl[2 \!-\! \epsilon(t)\Bigr] 
- 6\dot{\epsilon}(t) H^3(t)\Biggr\} dt \; .
\end{equation}
One finds $f(R)$ by integrating the relation for $f'(R)$ and using 
(\ref{tofR}),
\begin{equation}
f(R) = f(R_i) + \int_{t_i}^{t(R)} \!\! dR_0(t') \exp\Biggl[ 
\sqrt{\frac{16 \pi G}{3}} \, \varphi\Bigl( \widetilde{t}(t') \Bigr)
\Biggr] \; . 
\end{equation} 

\section{Comparing Analytic and Numerical Results}

The purpose of this section is to compare analytic and numerical 
results for Starobinsky inflation and another representative $f(R)$ 
model. We begin by explaining how the analytic results are derived.
Then the models are described and numerical results for their power
spectra are given. The section closes by comparing with various 
analytic approximations.

\subsection{How We Compute $\Delta^2_{\mathcal{R}}(k)$ and 
$\Delta^2_{h}(k)$}

We use the Hubble representation \cite{Liddle:1994dx} of the Einstein 
frame, in which one assumes that $\widetilde{a}(\widetilde{t}) \equiv 
\widetilde{a}_i e^{\widetilde{n}}$, $\widetilde{H}(\widetilde{t})$ and 
$\widetilde{\epsilon}(\widetilde{t})$ are known, or can be generated
numerically. Because the Einstein frame is a scalar potential model we
represent the power spectra the same as expressions 
(\ref{fullDR}-\ref{fullDh}) but using the Einstein frame geometry,
\begin{eqnarray}
\Delta^2_{h}(k) & = & \frac{16}{\pi} \, G \widetilde{H}^2(\widetilde{t}_k) 
\!\times\! C\Bigl( \widetilde{\epsilon}(\widetilde{t}_k)\Bigr) \!\times\! 
\widetilde{\mathcal{C}}(k) \; , \label{fulltildeDh} \\
\Delta^2_{\mathcal{R}}(k) & = & \frac{G \widetilde{H}^2(\widetilde{t}_k)}{
\pi \widetilde{\epsilon}(\widetilde{t}_k)} \!\times\! C\Bigl( 
\widetilde{\epsilon}(\widetilde{t}_k)\Bigr) \!\times\! 
\widetilde{\mathcal{S}}(k) \; . \label{fulltildeDR}
\end{eqnarray}
Here the slow roll correction factor $C(\epsilon)$ was defined in
(\ref{Cdef}). Of course the terms involving $\widetilde{H}(\widetilde{t}_k)$ 
and $\widetilde{\epsilon}(\widetilde{t}_k)$ are clear enough so it is the 
nonlocal correction factors, $\widetilde{\mathcal{C}}(k)$ and
$\widetilde{\mathcal{S}}(k)$ which require explanation.

Our technique for determining the nonlocal correction factors is based 
on nonlinear evolution equations \cite{Romania:2012tb} for the 
norm-squared mode functions $\widetilde{M}(\widetilde{t},k) \equiv 
\vert \widetilde{u}(\widetilde{t},k) \vert^2$ and $\widetilde{N}(
\widetilde{t},k) \equiv \vert \widetilde{v}(\widetilde{t},k) \vert^2$ 
which appear in expressions (\ref{powerhtilde}) and (\ref{powerRtilde}) 
for the power spectra. We then factor out the instantaneously constant 
$\widetilde{\epsilon}$ solutions and express the residuals in terms
of the number of e-foldings $\widetilde{n}$ since the beginning of
inflation \cite{Brooker:2015iya,Brooker:2016xkx},
\begin{equation}
\widetilde{M}(\widetilde{t},k) \equiv \widetilde{M}_0(\widetilde{t},k)
\times \exp\Bigl[ -\frac12 \widetilde{h}(\widetilde{n},k)\Bigr] 
\; , \; \widetilde{N}(\widetilde{t},k) \equiv \frac{
\widetilde{M}_0(\widetilde{t},k)}{\widetilde{\epsilon}(\widetilde{t},k)}
\times \exp\Bigl[ -\frac12 \widetilde{g}(\widetilde{n},k)\Bigr]
\; ,
\end{equation}
where the instantaneously constant $\widetilde{\epsilon}$ solution 
involves a Hankel function,
\begin{equation}
\widetilde{M}(\widetilde{t},k) \equiv \frac{\pi}{[1 \!-\! 
\widetilde{\epsilon}(\widetilde{t})] \widetilde{H}(\widetilde{t})
\widetilde{a}^3(\widetilde{t})} \Biggl\vert H^{(1)}_{\widetilde{\nu}(
\widetilde{t})}\Biggl( \frac{k}{1 \!-\! \widetilde{\epsilon}(\widetilde{t})]
\widetilde{H}(\widetilde{t}) \widetilde{a}(\widetilde{t})}\Biggr)
\Biggr\vert^2 \; , \; \widetilde{\nu} \equiv \frac12 \Bigl( \frac{3 \!-\!
\widetilde{\epsilon}}{1 \!-\! \widetilde{\epsilon}}\Bigr) \; .
\end{equation}
The nonlocal correction factors come from the late time forms of the
residuals $\widetilde{h}(\widetilde{n},k)$ and 
$\widetilde{g}(\widetilde{n},k)$,
\begin{eqnarray}
\widetilde{\mathcal{C}}(k) & \!\!\!=\!\!\! & 
\lim_{\widetilde{t} \gg \widetilde{t}_k} 
\Biggl[ \frac{\widetilde{a}(\widetilde{t})}{\widetilde{a}(\widetilde{t}_k)}
\Biggr]^{\frac{2 \widetilde{\epsilon}(\widetilde{t})}{1 - 
\widetilde{\epsilon}(\widetilde{t})}} \times
\Biggl[ \frac{\widetilde{H}(\widetilde{t})}{\widetilde{H}(\widetilde{t}_k)}
\Biggr]^{\frac{2}{1 - \widetilde{\epsilon}(\widetilde{t})}} \times 
\frac{C(\widetilde{\epsilon}(\widetilde{t}))}{C(\widetilde{\epsilon}(
\widetilde{t}_k)} \times \exp\Bigl[-\frac12 \widetilde{h}(\widetilde{n},k)
\Bigr] \; , \quad \\
\widetilde{\mathcal{S}}(k) & \!\!\!=\!\!\! & 
\lim_{\widetilde{t} \gg \widetilde{t}_k} 
\Biggl[ \frac{\widetilde{a}(\widetilde{t})}{\widetilde{a}(\widetilde{t}_k)}
\Biggr]^{\frac{2 \widetilde{\epsilon}(\widetilde{t})}{1 - 
\widetilde{\epsilon}(\widetilde{t})}} \times
\Biggl[ \frac{\widetilde{H}(\widetilde{t})}{\widetilde{H}(\widetilde{t}_k)}
\Biggr]^{\frac{2}{1 - \widetilde{\epsilon}(\widetilde{t})}} \nonumber \\
& & \hspace{4.5cm} \times 
\frac{C(\widetilde{\epsilon}(\widetilde{t}))}{C(\widetilde{\epsilon}(
\widetilde{t}_k)} \times \frac{\widetilde{\epsilon}(\widetilde{t}_k)}{
\widetilde{\epsilon}(\widetilde{t})} \times \exp\Bigl[-\frac12 
\widetilde{g}(\widetilde{n},k) \Bigr] \; . \quad 
\end{eqnarray}

The residuals are damped, driven oscillators with small nonlinearities
\cite{Brooker:2015iya,Brooker:2016xkx},
\begin{eqnarray}
\widetilde{h}'' - \frac{\widetilde{\omega}'}{\widetilde{\omega}} 
\widetilde{h}' + \widetilde{\omega}^2 \widetilde{h} & = & \widetilde{S} 
+ \frac14 \Bigl( \widetilde{h}'\Bigr)^2 + \widetilde{\omega}^2 
\Bigl[1 \!+\! \widetilde{h} \!-\! e^{\widetilde{h}}\Bigr] \; , 
\label{heqn} \\
\widetilde{g}'' - \frac{\widetilde{\omega}'}{\widetilde{\omega}} 
\widetilde{g}' + \widetilde{\omega}^2 \widetilde{g} & = & \widetilde{S}
+ \Delta \widetilde{S} + \frac14 \Bigl( \widetilde{g}'\Bigr)^2 + 
\widetilde{\omega}^2 \Bigl[1 \!+\! \widetilde{g} \!-\! 
e^{\widetilde{g}}\Bigr] \; . \label{geqn}
\end{eqnarray}
Here and henceforth a prime denotes differentiation with respect to
$\widetilde{n}$. It is remarkable that both the tensor and scalar
residual have the same frequency,
\begin{equation}
\widetilde{\omega}(\widetilde{n},k) \equiv \frac1{\widetilde{H}(\widetilde{n})
\widetilde{a}^3(\widetilde{t}) \widetilde{M}_0(\widetilde{t},k)} \; .
\end{equation}
The source for the tensor residual vanishes for constant 
$\widetilde{\epsilon}$ \cite{Brooker:2015iya} and is typically small,
\begin{equation}
\widetilde{S}(\widetilde{n},k) \equiv \frac{4 k^2}{\widetilde{H}^2 
\widetilde{a}^2} - \widetilde{\omega}^2 + 2 \Biggl[ 
\frac{\widetilde{M}_0''}{\widetilde{M}_0} -\frac12 \Bigl( 
\frac{\widetilde{M}_0'}{\widetilde{M}_0}\Bigr)^2 + (3 \!-\! 
\widetilde{\epsilon}) \frac{\widetilde{M}_0'}{\widetilde{M}_0}
\Biggr] \; .
\end{equation}
In contrast, the extra source for the scalar residual can be large if 
the potential has features \cite{Brooker:2016xkx},
\begin{equation}
\Delta \widetilde{S}(\widetilde{n}) \equiv -2 \Biggl[ 
\frac{\widetilde{\epsilon}''}{\widetilde{\epsilon}} -\frac12
\Bigl( \frac{\widetilde{\epsilon}'}{\widetilde{\epsilon}}\Bigr)^2
+ (3 \!-\! \widetilde{\epsilon}) \frac{\widetilde{\epsilon}'}{
\widetilde{\epsilon}}\Biggr] \; .
\end{equation}

Another remarkable fact is that the linear differential operators on 
the left hand side of (\ref{heqn}-\ref{geqn}) possess a Green's function 
which is known analytically for an {\it arbitrary} inflationary
expansion history \cite{Brooker:2015iya,Brooker:2016xkx},
\begin{equation}
\widetilde{G}(\widetilde{n};\widetilde{m}) = 
\frac{\theta(\widetilde{n} \!-\! \widetilde{m})}{
\widetilde{\omega}(\widetilde{m},k)} \, \sin\Biggl[ 
\int_{\widetilde{m}}^{\widetilde{n}} \!\! d\ell \,
\widetilde{\omega}(\ell,k)\Biggr] \; .
\end{equation}
This means we can express both residuals analytically as series 
expansions $\widetilde{h} = \widetilde{h}_1 + \widetilde{h}_2 + \dots$ 
and $\widetilde{g} = \widetilde{g}_1 + \widetilde{g}_2 + \dots$, whose
first two terms are,
\begin{eqnarray}
\widetilde{h}_1(\widetilde{n},k) & = & \int_{0}^{\widetilde{n}} \!\!
d\widetilde{m} \, \widetilde{G}(\widetilde{n};\widetilde{m}) 
\widetilde{S}(\widetilde{m},k) \; , \label{h1} \\
\widetilde{g}_1(\widetilde{n},k) & = & \int_{0}^{\widetilde{n}} \!\!
d\widetilde{m} \, \widetilde{G}(\widetilde{n};\widetilde{m}) \Bigl[
\widetilde{S}(\widetilde{m},k) + \Delta \widetilde{S}(\widetilde{m})
\Bigr] \; , \label{g1} \\
\widetilde{h}_2(\widetilde{n},k) & = & \int_{0}^{\widetilde{n}} \!\!
d\widetilde{m} \, \widetilde{G}(\widetilde{n};\widetilde{m}) 
\Biggl[ \frac14 \widetilde{h}_1^{\prime 2} (\widetilde{m},k)
- \frac12 \widetilde{\omega}^2(\widetilde{m},k) 
\widetilde{h}_1^2(\widetilde{m},k)\Biggr] \; , \label{h2} \\ 
\widetilde{g}_2(\widetilde{n},k) & = & \int_0^{\widetilde{n}} \!\! 
d\widetilde{m} \, \widetilde{G}(\widetilde{n};\widetilde{m}) 
\Biggl[ \frac14 \widetilde{g}_1^{\prime 2} (\widetilde{m},k)
- \frac12 \widetilde{\omega}^2(\widetilde{m},k) 
\widetilde{g}_1^2(\widetilde{m},k)\Biggr] \; . \label{g2}
\end{eqnarray}
The higher terms --- $\widetilde{h}_2(\widetilde{n},k)$,
$\widetilde{g}_2(\widetilde{n},k)$ and so on --- are only necessary
if the residuals or their derivatives become order one or larger.

\begin{figure}[ht]
\includegraphics[width=10cm,height=8cm]{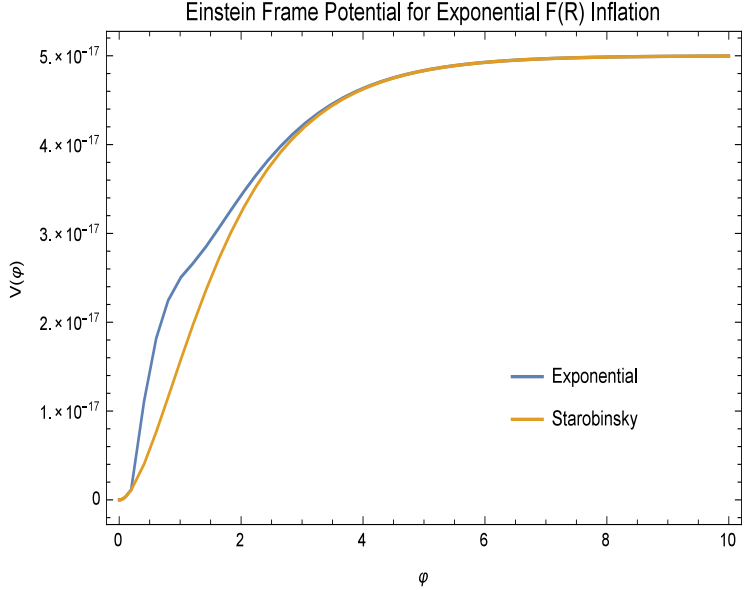}
\caption{Comparison of the potentials $V(\varphi)$ for Starobinsky
inflation (yellow) and the exponential model (blue).}
\label{potentials}
\end{figure}

Although expressions (\ref{h1}-\ref{g2}) involve integrations over
the entire range of e-foldings from the beginning of inflation, the
only net contributions come from the few e-foldings around first
horizon crossing. The reason nothing happens before is that the
frequency term is so large at early times,
\begin{equation}
{\rm Early\ Times:} \qquad \widetilde{\omega}^2(\widetilde{n},k) = 
\Bigl( \frac{2k}{\widetilde{H} \widetilde{a}}\Bigr)^2 \Biggl[ 1 + O\Bigl( 
\frac{\widetilde{H}^2 \widetilde{a}^2}{k^2}\Bigr) \Biggr] \; .
\end{equation}
This means that the early time form of the scalar residual is small,
the tensor residual is very small, and both are local 
\cite{Brooker:2015iya,Brooker:2016xkx},
\begin{eqnarray}
{\rm Early\ Times:} \; \widetilde{g}(\widetilde{n},k) & = & 
\Delta \widehat{S}(\widetilde{n}) \!\times\! \Bigl( \frac{\widetilde{H}
\widetilde{a}}{2 k}\Bigr)^2 + O\Bigl( \frac{\widetilde{H}^4 
\widetilde{a}^4}{k^4}\Bigr) \; , \\
{\rm Early\ Times:} \; \widetilde{h}(\widetilde{n},k) & = & -4
\Bigl[\widetilde{\epsilon}'' + (9 \!-\! 7 \widetilde{\epsilon})
\widetilde{\epsilon}'\Bigr] \!\times\! \Bigl( \frac{\widetilde{H}
\widetilde{a}}{2 k}\Bigr)^4 + O\Bigl( \frac{\widetilde{H}^6 
\widetilde{a}^6}{k^6}\Bigr) \; . \qquad 
\end{eqnarray}
Shortly after first horizon crossing the frequency drops to zero,
\begin{equation}
{\rm Late\ Times:} \; \widetilde{\omega}^2(\widetilde{n},k) =
\Bigl( \frac{2 k}{\widetilde{H} \widetilde{a}}\Bigr)^{\frac{6 - 2
\widetilde{\epsilon}}{1 - \widetilde{\epsilon}}} \Biggl[ \frac{\pi^2}{
[4 (1 \!-\! \widetilde{\epsilon})]^{\frac4{1 -\widetilde{\epsilon}}}
\Gamma^4( \frac32 \!+\! \frac{\widetilde{\epsilon}}{1 - 
\widetilde{\epsilon}})} + O\Bigl( \frac{k^2}{\widetilde{H}^2 
\widetilde{a}^2}\Bigr) \Biggr] \; .
\end{equation}
Although the residuals $\widetilde{h}(\widetilde{n},k)$ and 
$\widetilde{g}(\widetilde{n},k)$ have some small late time dependence
due to continued evolution of $\widetilde{\epsilon}(\widetilde{t})$,
the full solutions $\widetilde{M}(\widetilde{t},k)$ and
$\widetilde{N}(\widetilde{t},k)$ freeze in to constant values less
than two e-foldings after horizon crossing. 

\subsection{The Two Models}

\begin{figure}[ht]
\includegraphics[width=10cm,height=8cm]{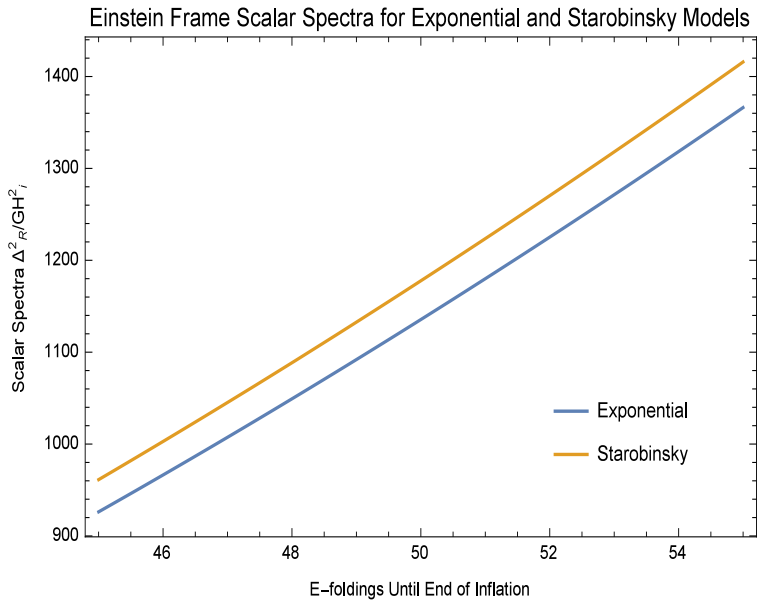}
\caption{Comparison of the scalar power spectrum $\Delta^2_{\mathcal{R}}(k)$
for Starobinsky inflation (yellow) and the exponential model (blue).
Both are displayed as a function of $\widetilde{N}$, the number of
Einstein frame e-foldings before the end of inflation at which horizon 
crossing occurs.}
\label{scalarEinstein}
\end{figure}

\begin{figure}[ht]
\includegraphics[width=6.0cm,height=4.8cm]{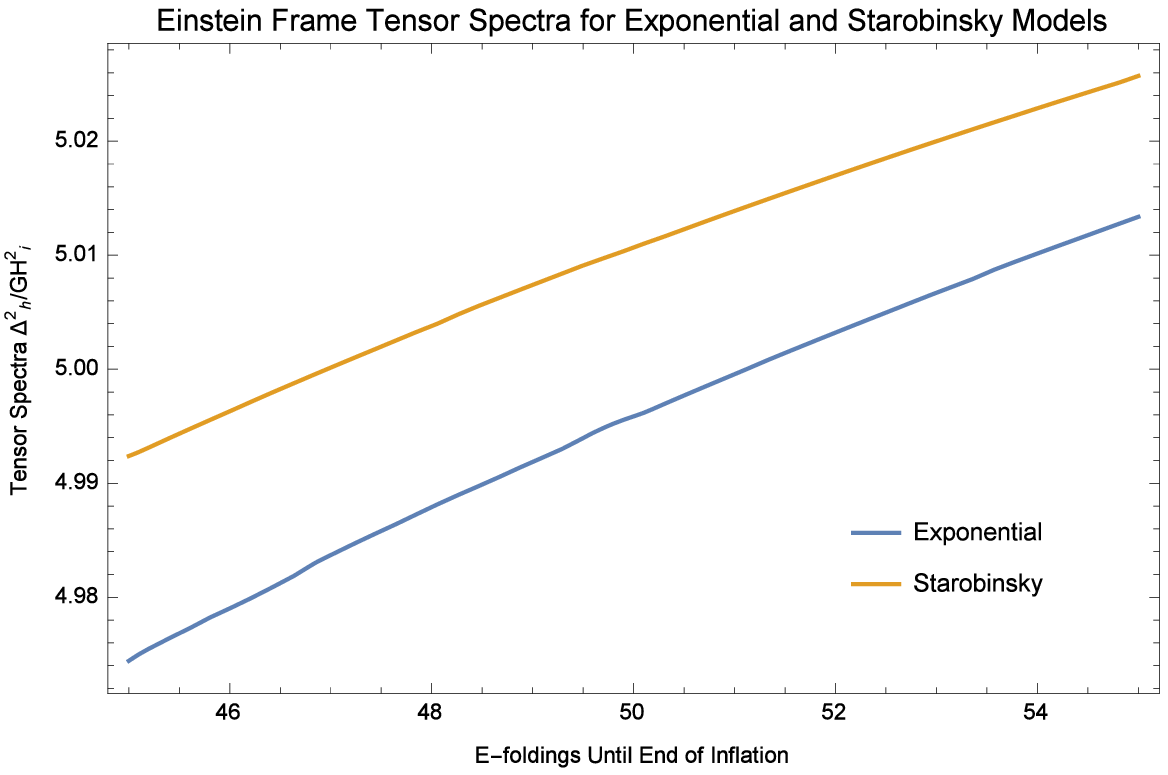}
\hspace{1cm}
\includegraphics[width=6.0cm,height=4.8cm]{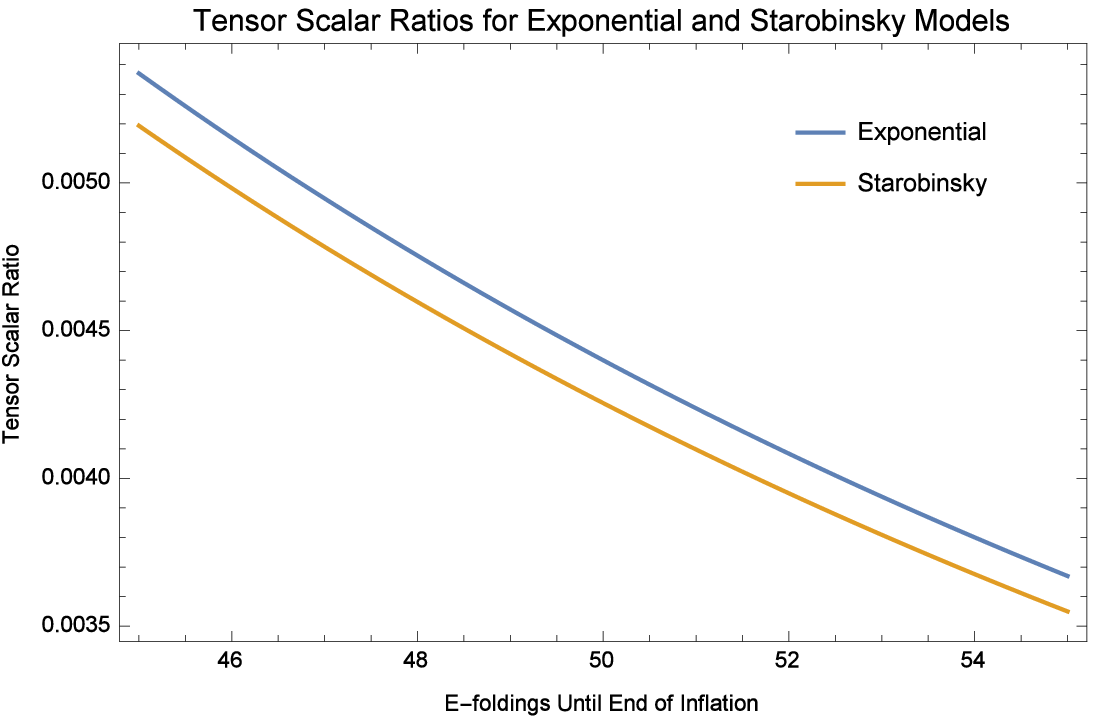}
\caption{Comparison of the tensor power spectrum $\Delta^2_{h}(k)$ 
(left) and the tensor-to-scalar ratio $r(k)$ (right) for Starobinsky 
inflation (yellow) and the exponential model (blue). All results are 
displayed as a function of $\widetilde{N}$, the number of Einstein 
frame e-foldings before the end of inflation at which horizon 
crossing occurs.}
\label{tensorEinstein}
\end{figure}

We studied two models, both of which take the form (\ref{LJ}). The
first was Starobinsky inflation (\ref{AAS}), with the parameter 
and initial conditions chosen as,
\begin{equation}
M = 10^{-5} \qquad , \qquad \epsilon_{i} = 0.00221 \qquad , \qquad 
G H^2_i = 7.55 \!\times\! 10^{-9} \; . \label{staroparams}
\end{equation}
We also studied a model which has been proposed to describe cosmology 
from inflation to the current phase of acceleration 
\cite{Elizalde:2010ts},
\begin{equation}
f(R) = R - \Lambda \Biggl[1 \!-\! \exp\Bigl[-\Bigl( \frac{R}{2 
\Lambda}\Bigr)^{4} \Bigr] \Biggr] + \frac{R^2}{ 4 \Lambda} \; .
\label{expmodel}
\end{equation}
The parameter and initial conditions were chosen as,
\begin{equation}
G \Lambda = 10^{-16} \qquad , \qquad \epsilon_i=0.00501 \qquad , \qquad
G H^2_i = 2.22 \!\times\! 10^{-15} \; . \label{expparams}
\end{equation}
Despite the different functions $f(R)$ between (\ref{AAS}) and 
(\ref{expmodel}), the two models are quite similar as far as inflation
is concerned. This shows up clearly from Figure~\ref{potentials} which 
gives their potentials. Although there are some significant differences
for low potential, inflation is governed by the behavior for large
potential, which is almost identical.

\subsection{Power Spectra of the Two Models}

\begin{figure}[ht]
\includegraphics[width=4.0cm,height=3.2cm]{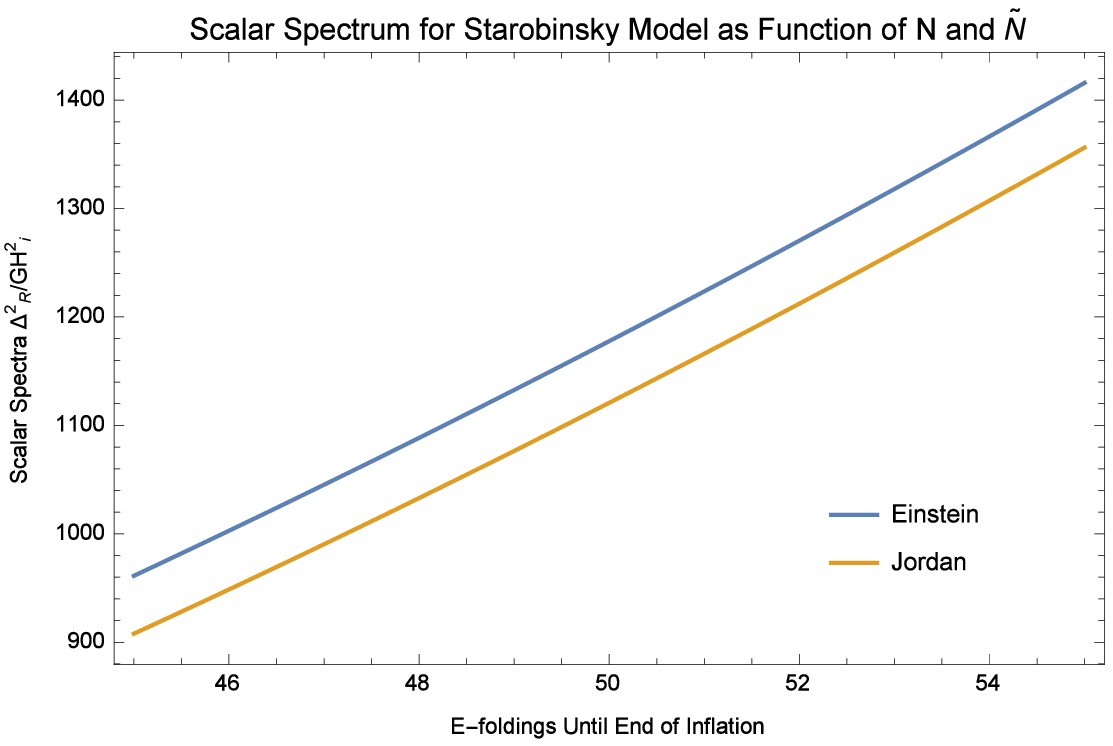}
\hspace{.5cm}
\includegraphics[width=4.0cm,height=3.2cm]{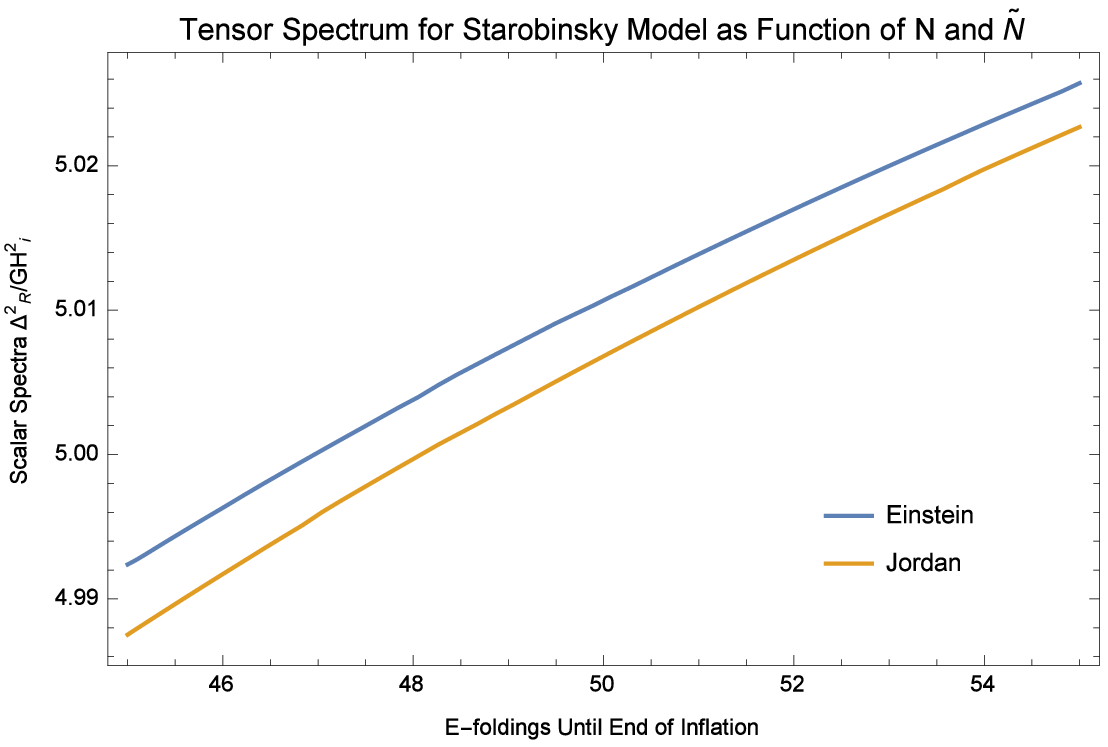}
\hspace{.5cm}
\includegraphics[width=4.0cm,height=3.2cm]{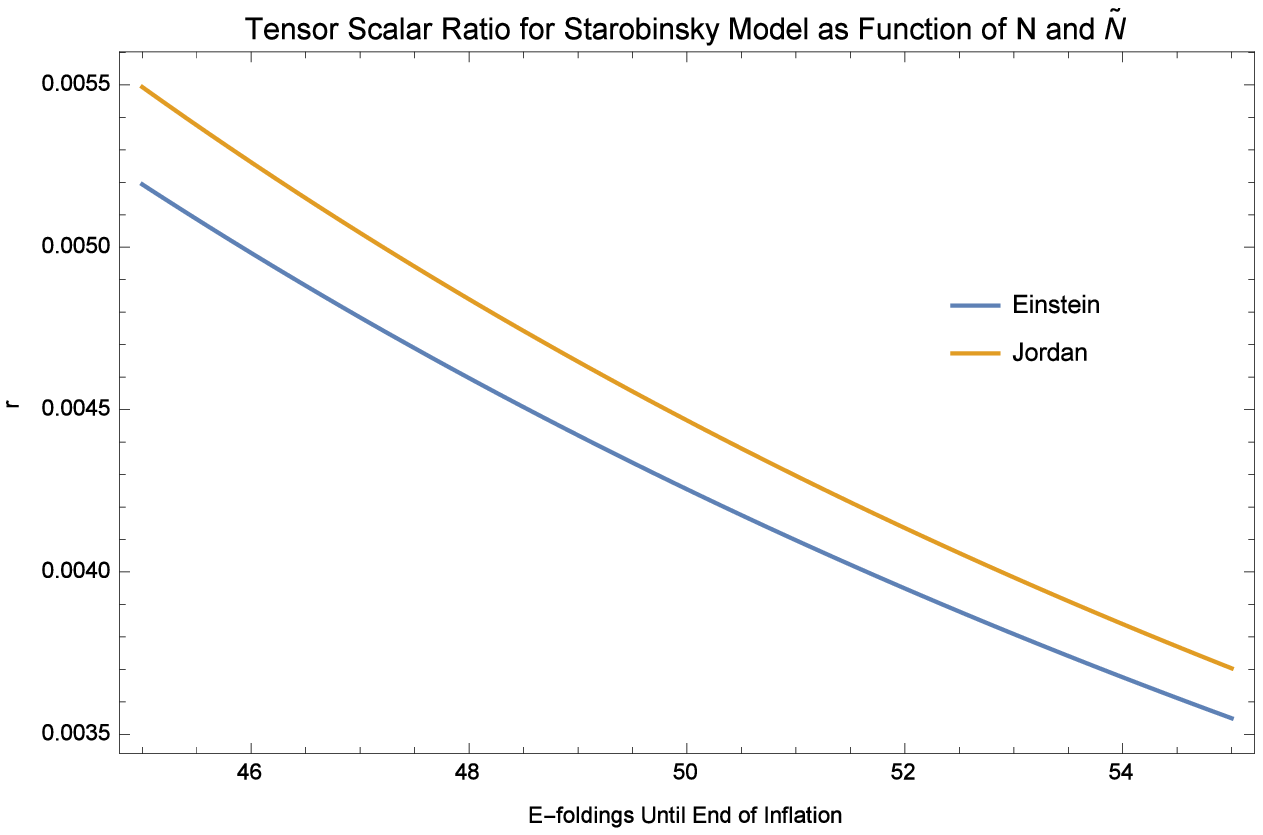}
\caption{The various spectra --- $\Delta^2_{\mathcal{R}}(k)$
(left), $\Delta^2_{h}(k)$ (middle) and $r(k)$ (right) --- 
for Starobinsky inflation, as functions of the number of 
e-foldings from first horizon crossing until the end of inflation. 
For the yellow plots the $x$ axes give $N$, the number of e-foldings 
in the Jordan frame, whereas the $x$ axes of the blue plots give 
$\widetilde{N}$, the number of e-foldings in the Einstein frame.}
\label{R2EvsJ}
\end{figure}

\begin{figure}[ht]
\includegraphics[width=4.0cm,height=3.2cm]{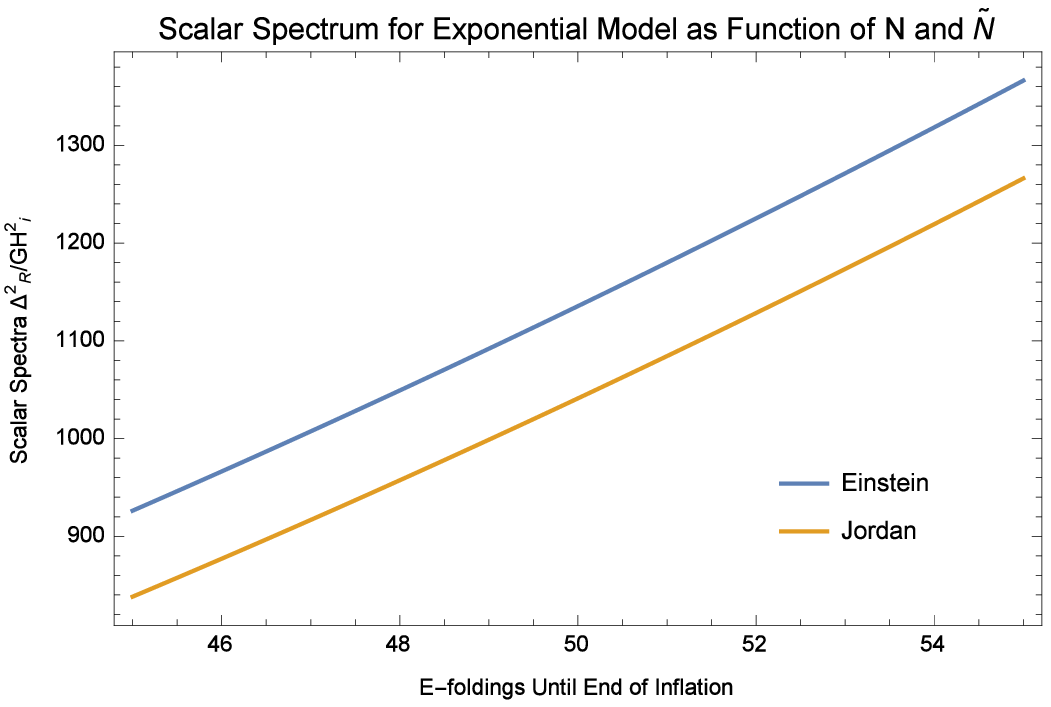}
\hspace{.5cm}
\includegraphics[width=4.0cm,height=3.2cm]{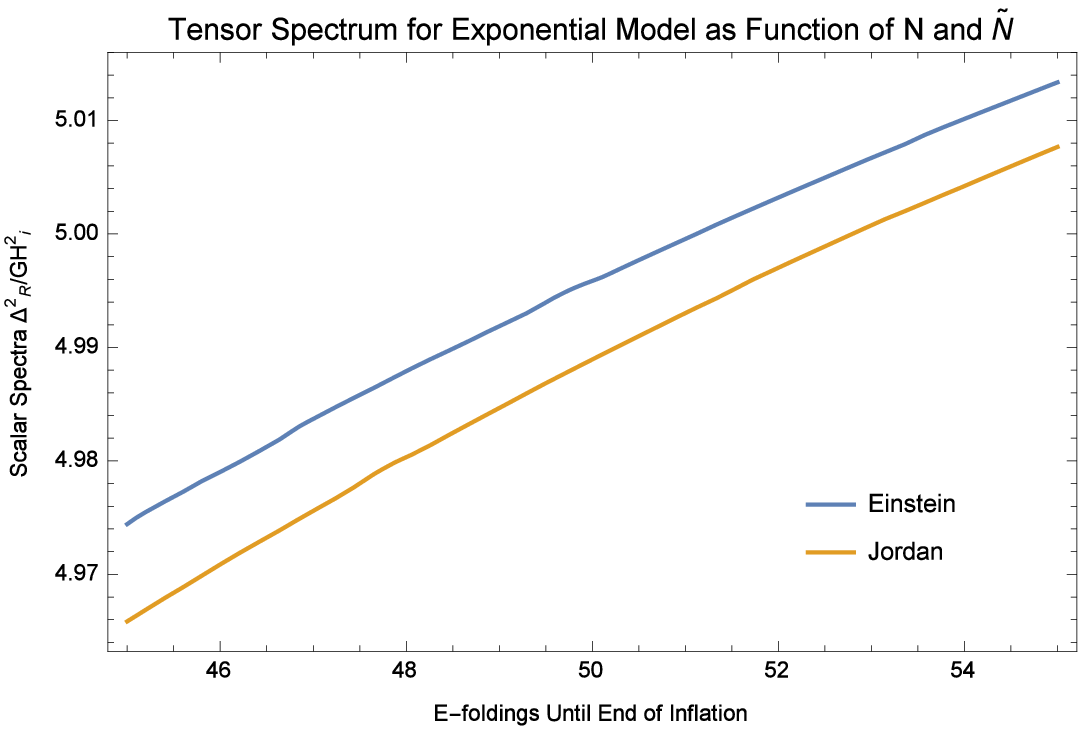}
\hspace{.5cm}
\includegraphics[width=4.0cm,height=3.2cm]{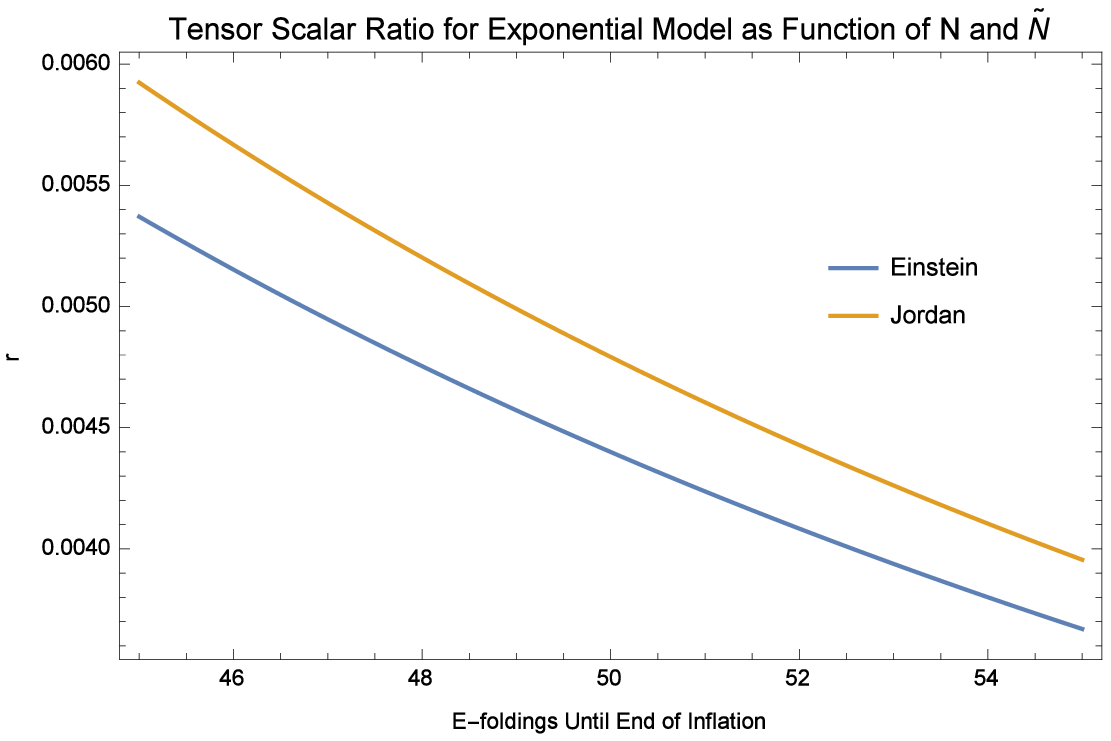}
\caption{The various spectra --- $\Delta^2_{\mathcal{R}}(k)$
(left), $\Delta^2_{h}(k)$ (middle) and $r(k)$ (right) --- 
for the exponential model, as functions of the number of 
e-foldings from first horizon crossing until the end of inflation. 
For the yellow plots the $x$ axes give $N$, the number of e-foldings 
in the Jordan frame, whereas the $x$ axes of the blue plots give 
$\widetilde{N}$, the number of e-foldings in the Einstein frame.}
\label{ExpEvsJ}
\end{figure}

We numerically simulated each model exactly. Figure~\ref{scalarEinstein} 
shows that the scalar power spectrum of the Starobinsky inflation is 
slightly larger than for exponential model, although both have roughly 
the same shape. From Figure~\ref{tensorEinstein} we see that the tensor 
power spectrum of Starobinsky inflation slight exceeds that of the 
exponential model. However, the difference is so slight that the 
tensor-to-scalar ratio of the exponential model exceeds that of 
Starobinsky inflation.

\begin{figure}[ht]
\includegraphics[width=4.0cm,height=3.2cm]{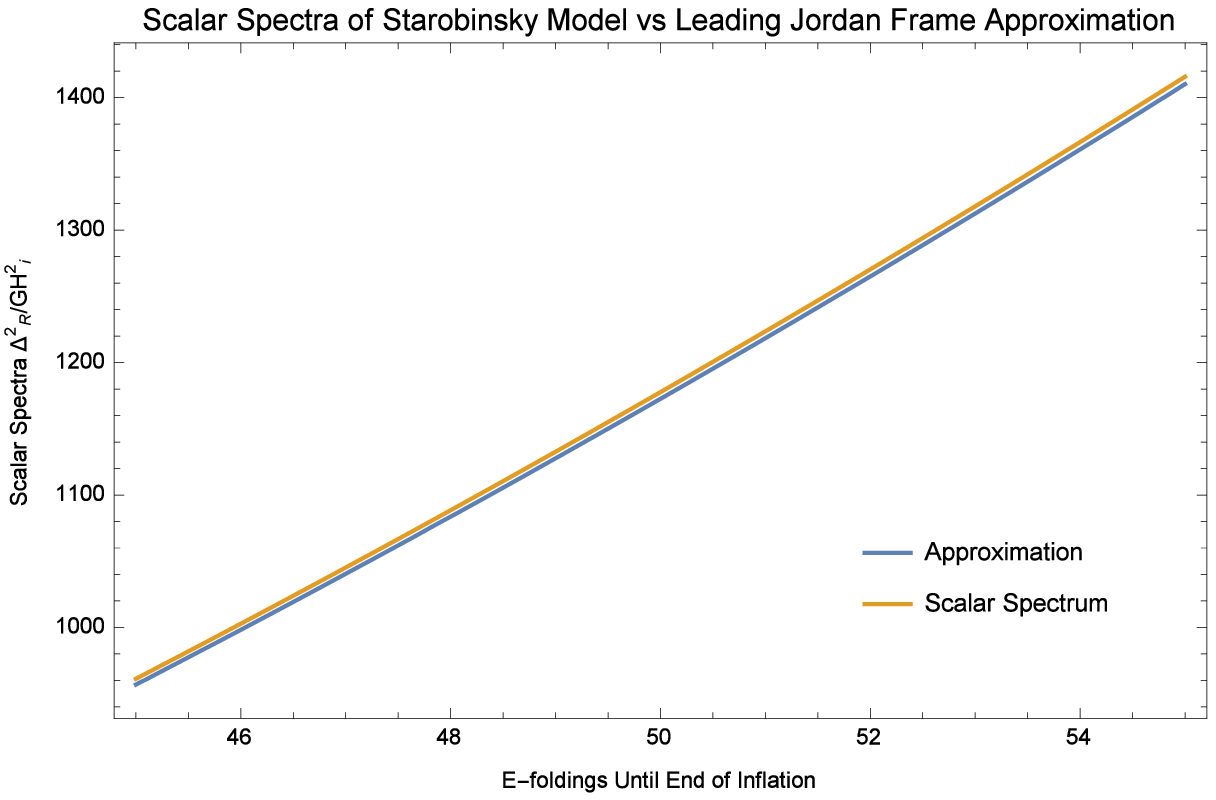}
\hspace{.5cm}
\includegraphics[width=4.0cm,height=3.2cm]{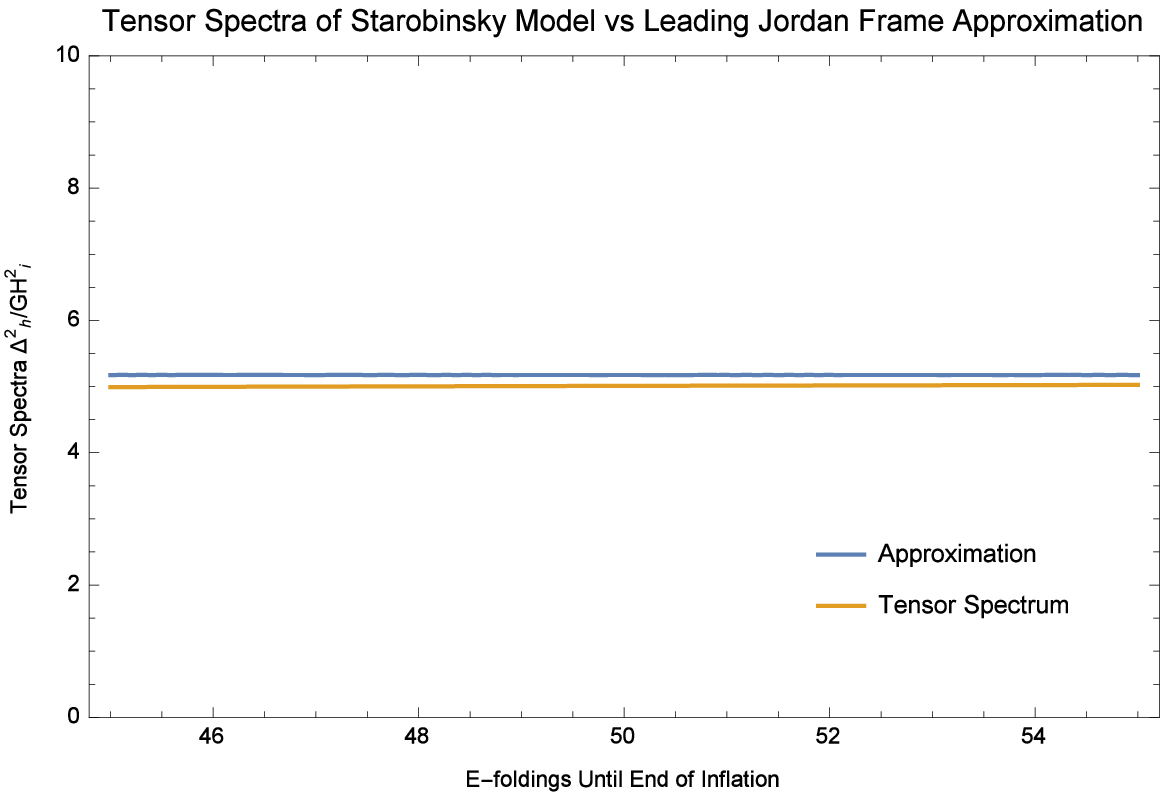}
\hspace{.5cm}
\includegraphics[width=4.0cm,height=3.2cm]{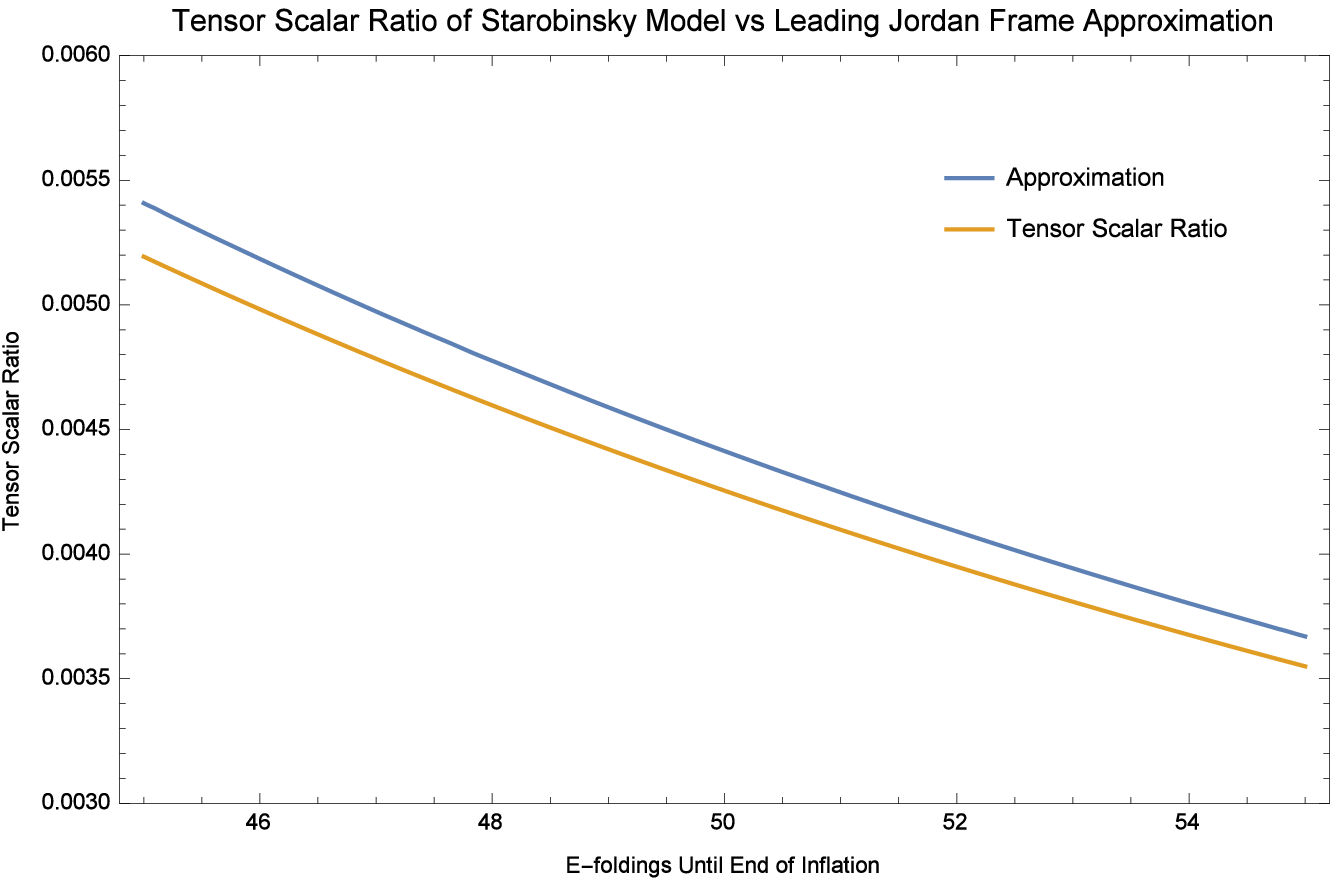}
\caption{Comparison of the exact results (yellow) with the leading
slow roll approximation (blue) for Starobinsky inflation. The left
graph shows the scalar power spectrum (\ref{DR}), the middle graph
shows the tensor power spectrum (\ref{Dh}), and the right graph
show the tensor-to-scalar ratio (\ref{r}).}
\label{LeadingApprox}
\end{figure}

\begin{figure}[ht]
\includegraphics[width=6.0cm,height=4.8cm]{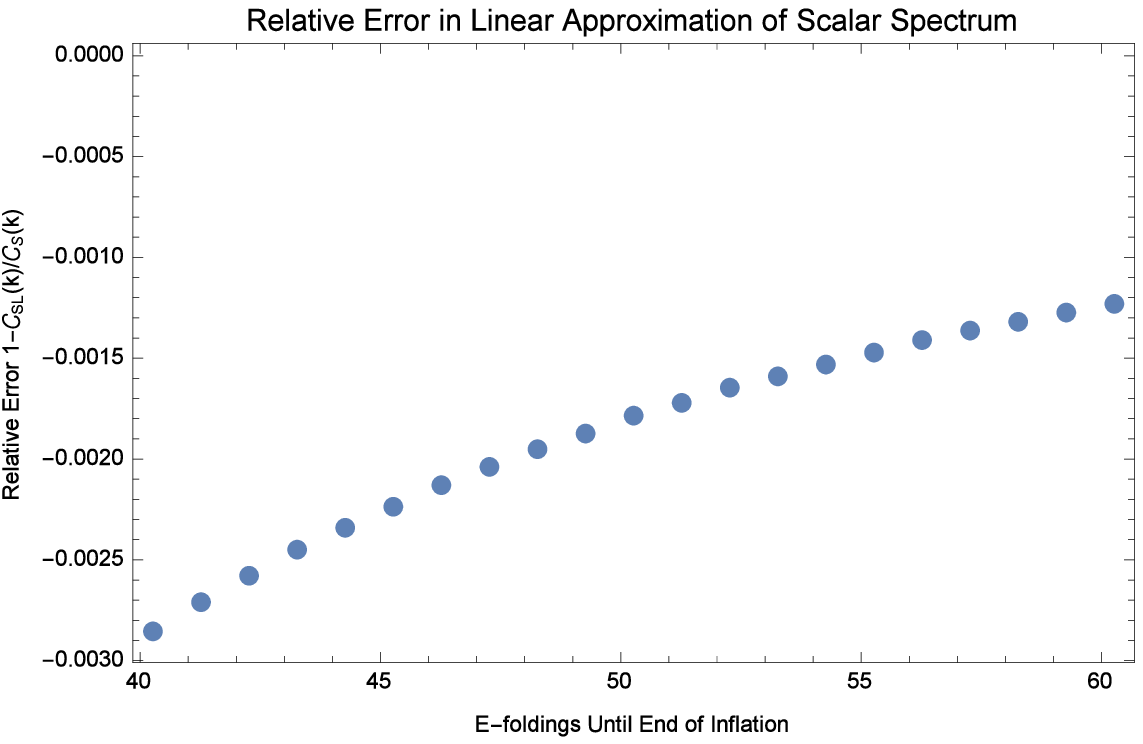}
\hspace{1cm}
\includegraphics[width=6.0cm,height=4.8cm]{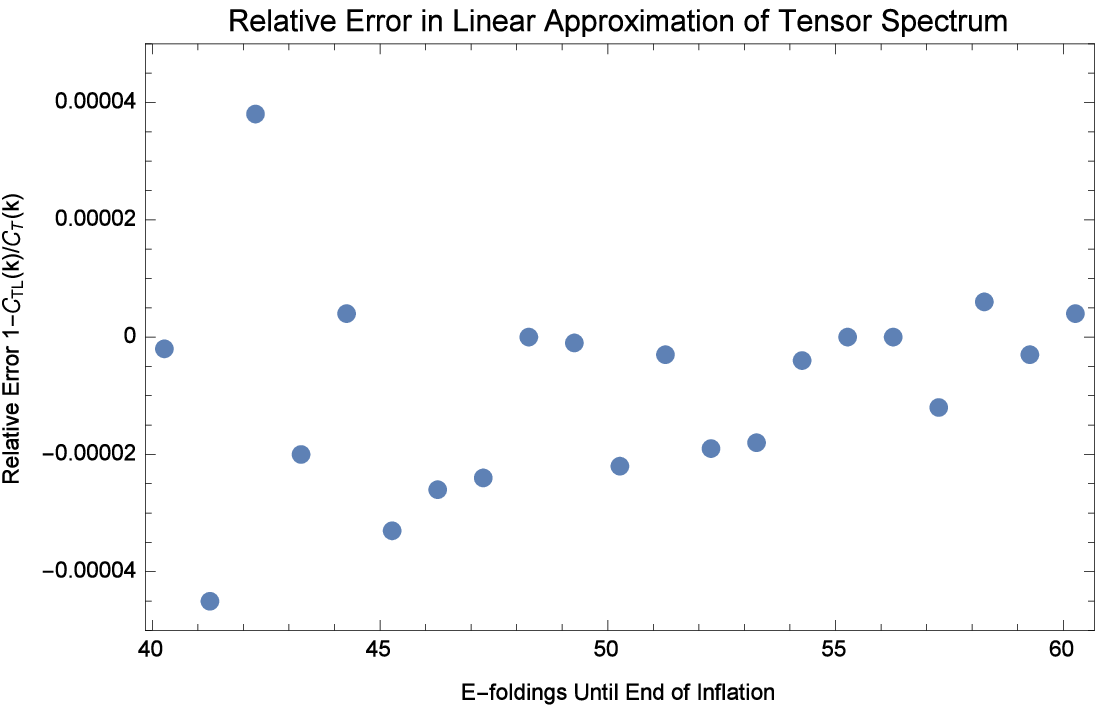}
\caption{Fractional error of our linearized approximation to the
scalar (left) and tensor (right) power spectra for Starobinsky 
inflation.}
\label{LinearizedErr}
\end{figure}

Figure~\ref{R2EvsJ} displays the spectra of the Starobinsky model
as functions of the number of e-foldings $N$ to the end of inflation
in the Jordan frame, and the number of e-foldings $\widetilde{N}$ to the
end of inflation in the Einstein frame. In each case, features at $N$
appear to be displaced to $\widetilde{N} \simeq N -\frac12 \ln(1 + 2N)$,
in agreement with equation(\ref{NtoNtilde}). Figure~\ref{ExpEvsJ} gives 
the relation between $N$ and $\widetilde{N}$ for the exponential model.

\subsection{Comparison with Analytic Results}

A major point of this paper has been to develop good analytic 
approximations for how the power spectra of $f(R)$ models depend
functionally upon the geometry. For all the spectra, and for both 
models, the leading slow roll approximations are pretty accurate,
\begin{eqnarray}
\Delta^2_{\mathcal{R}}(k) & \simeq & \frac{G \widetilde{H}^2(
\widetilde{t}_k)}{\pi \widetilde{\epsilon}(\widetilde{t}_k)}
\simeq \frac{G H^2(t_k)}{2 \pi \epsilon(t_k)} 
\; , \label{leadDR} \\
\Delta^2_{h}(k) & \simeq & \frac{16}{\pi} G \widetilde{H}^2(
\widetilde{t}_k) \simeq \frac{24}{\pi} \, G H^2(t_k) 
\epsilon(t_k) \; , \label{leadDh} \\
r(k) & \simeq & 16 \widetilde{\epsilon}(\widetilde{t}_k) 
\simeq 48 \epsilon^2(t_k) \; . \label{leadDr}
\end{eqnarray}
Figure~\ref{LeadingApprox} shows this for Starobinsky inflation.

Including the slow roll corrections, and just the linearized
approximations for $\mathcal{S}(k)$ and $\mathcal{C}$, makes the 
agreement essentially perfect. Figure~\ref{LinearizedErr} shows that 
the relative error of the scalar power spectrum is less than $0.3\%$ 
for Starobinsky inflation. The relative error for the tensor power 
spectrum is actually at the $0.002\%$ accuracy of our numerical 
simulation.

\section{Discussion}

We have developed a good functional form for the primordial power 
spectra of $f(R)$ inflation, after discussing (in section 2) the 
relation between Jordan and Einstein frames. When the Einstein frame 
potential lacks features, the leading slow roll results 
(\ref{leadDR}-\ref{leadDr}) are accurate. This is shown for Starobinsky 
inflation by Figure~\ref{LeadingApprox}. (An $f(R)$ model will agree 
with Starobinsky inflation if the parameter $X(t)$ of equation (49) 
obeys $X(t) \simeq -\epsilon(t)$.) When features are present (for which 
there continues to be observational support \cite{Hazra:2016fkm}), one 
gets essentially perfect agreement by using just the first two terms of 
the nonlocal correction factors (\ref{h1}-\ref{g2}) in expressions
(\ref{fulltildeDh}-\ref{fulltildeDR}) \cite{Brooker:2016xkx}.

One cannot distinguish $f(R)$ models from scalar potential models with 
just the power spectra. In section 3 we showed how the same data could be 
used to reconstruct either kind of model. Even for de Sitter-like models 
this changes if one has information about what the wave number ``$k$'' 
means in terms of other scales. There is a shift of 2-3 e-foldings between 
the same feature of the scalar potential reconstruction and the $f(R)$ 
reconstruction, with the scalar potential model feature appearing nearer
to the end of inflation. One can see this from Figures~\ref{R2EvsJ} and
\ref{ExpEvsJ}.

Finally, we mention that an interesting and very topical application
of this formalism is perturbations for Higgs inflation 
\cite{Bezrukov:2007ep,Bezrukov:2010jz}. More generally, scalar models 
with a nonminimal coupling involve similar conformal transformations 
between Jordan and Einstein frames.

\newpage

\centerline{\bf Acknowledgements}

We are grateful for conversations and correspondence with J. Garcia-Bellido
and M. Sasaki. This work was partially supported by MINECO (Spain) Project 
FIS2013-44881-P; by the CSIC I-LINK1019 Project; by a travel grant from the 
University of Florida International Center, College of Liberal Arts and 
Sciences, Graduate School and Office of the Provost; by NSF grant 
PHY-1506513; and by the Institute for Fundamental Theory at the University 
of Florida.

\end{document}